%% file: aea_protostars_h2co.tex
\documentclass{aa} 

\usepackage{graphicx,natbib,rotating,amssymb,url,txfonts,hyperref} 
\bibpunct{(}{)}{;}{a}{}{,}

\graphicspath{{figures/}}
 
\newcommand{\citenp}[1]{\citeauthor{#1} \citeyear{#1}} 
 
\newcommand{\solarlum}{L$_{\sun}$\ }  
\newcommand{\solarmass}{M$_{\sun}$\ }

\newcommand{\Xin}{$X_\mathrm{in}$\ } 
\newcommand{\Xout}{$X_\mathrm{out}$\ }
\renewcommand{\object}{}
  
\begin{document} 
 
\date{\today} 
 
\title{The H$_2$CO abundance in the inner warm regions 
of low mass protostellar envelopes}

\titlerunning{The H$_2$CO abundance in low mass protostars}
 
\author{S. Maret\inst{1} \and C. Ceccarelli\inst{2} \and
  E. Caux\inst{1} \and A.G.G.M. Tielens\inst{3} \and J.K.  J\o
  rgensen\inst{4} \and E. van Dishoeck\inst{4} \and A. Bacmann\inst{5}
  \and A.Castets\inst{6} \and B. Lefloch\inst{2} \and
  L. Loinard\inst{7} \and B. Parise\inst{1} \and
  F. L. Sch\"oier\inst{4}}
 
\institute{Centre d'Etude Spatiale des Rayonnements, CESR/CNRS-UPS, BP
   4346, F-31028 Toulouse Cedex 04, France \and Laboratoire
   d'Astrophysique, Observatoire de Grenoble, B.P. 53, F-38041
   Grenoble Cedex 09, France \and Space Research Organization of the
   Netherlands, P.O. Box 800, 9700 AV Groningen, The Netherlands \and
   Leiden Observatory, P.O. Box 9513, NL-2300 RA Leiden, The
   Netherlands \and European Southern Observatory, Karl-Schwarzschild
   Str. 2, D-85748 Garching bei M\"unchen, Germany \and Observatoire
   de Bordeaux, BP 89, F-33270 Floirac, France \and Instituto de
   Astronom\'\i a, Universidad Nacional Aut\'onoma de M\'exico,
   Apartado Postal 72-3 (Xangari), 58089 Morelia, Michoac\'an, Mexico}
 
\offprints{S\'ebastien Maret, \email{sebastien.maret@cesr.fr}} 
 
\date{Received August 4, 2003 /Accepted October 9, 2003} 

\abstract{We present a survey of the formaldehyde emission in a sample
  of eight Class 0 protostars obtained with the IRAM and JCMT
  millimeter telescopes. The range of energies of the observed
  transitions allows us to probe the physical and chemical conditions
  across the protostellar envelopes. The data have been analyzed with
  three different methods with increasing level of sophistication.  We
  first analyze the observed emission in the LTE approximation, and
  derive rotational temperatures between 11 and 40 K, and column
  densities between 1 and $20 \times 10^{13}$ cm$^{-2}$. Second, we
  use a LVG code and derive higher kinetic temperatures, between 30
  and 90 K, consistent with subthermally populated levels and
  densities from 1 to $6 \times 10^5$ cm$^{-3}$.  The column densities
  from the LVG modeling are within a factor of 10 with respect to
  those derived in the LTE approximation. Finally, we analyze the
  observations based upon detailed models for the envelopes
  surrounding the protostars, using temperature and density profiles
  previously derived from continuum observations.  We approximate the
  formaldehyde abundance across the envelope with a jump function, the
  jump occurring when the dust temperature reaches 100 K, the
  evaporation temperature of the grain mantles.  The observed
  formaldehyde emission is well reproduced only if there is a jump of
  more than two orders of magnitude, in four sources.  In the
  remaining four sources the data are consistent with a formaldehyde
  abundance jump, but the evidence is more marginal ($\leq 2 ~
  \sigma$).  The inferred inner H$_2$CO abundance varies between $1
  \times 10^{-8}$ and $6 \times 10^{-6}$. The absolute values of the
  jump in the H$_2$CO abundance are uncertain by about one order of
  magnitude, because of the uncertainties in the density, ortho to
  para ratio, temperature and velocity profiles of the inner region,
  as well as the evaporation temperature of the ices. We discuss the
  implications of these jumps for our understanding of the origin and
  evolution of ices in low mass star forming regions.  Finally, we
  give predictions for the submillimeter H$_2$CO lines, which are
  particularly sensitive to the abundance jumps.

}

\maketitle
     
\section{Introduction} 
 
Low mass protostars form from dense fragments of molecular clouds.
During the pre-collapse and collapse phases, the physical and chemical
composition of the matter undergoes substantial, sometimes
spectacular, changes. From a chemical point of view, the pre-collapse
phase is marked by the freezing of molecules onto the grain mantles.
In the very inner parts of the pre-stellar condensations, molecules
have been observed to progressively disappear from the gas phase
\citep[e.g.][]{Tafalla02,Bergin02}.  The CO molecule, whose
condensation temperature is around 20 K, is the best studied species
both because it is the most abundant molecule after H$_2$, and because
of its important role in the gas thermal cooling. CO depletion of more
than a factor of ten has been observed in the centers of these
condensations \citep{Caselli98,Caselli02,Bacmann02}. This large CO
depletion is accompanied by a variety of changes in the molecular
composition; the most spectacular is the dramatic increase in the
molecular deuteration (up to eight orders of magnitude with respect to
the D/H elemental abundance) observed in formaldehyde
\citep{Bacmann03}.  The changes are recorded in the grain mantles,
where the pre-collapse gas will be progressively stored. When a
protostar is finally born, the dust cocoon warms up and the mantle
species evaporate into the gas phase, returning information from the
previous phase.
 
Most of the studies of the composition of the grain mantles have been
so far carried out towards massive protostars, because they have
strong enough IR continua against which the absorption of ices can be
observed \citep[e.g.][]{Gerakines99,Dartois99,Gibb00}.  The absorption
technique allows one to detect the most important constituents of the
grain mantles: H$_2$O, CO, CO$_2$, and sometime NH$_3$, CH$_3$OH and
H$_2$CO \citep{Schutte96, Keane01}. In much cases, the mantle
composition of low mass protostars has been directly observed. In
these cases, the observations have been carried out towards protostars
that possess a strong enough IR continuum \citep[e.g.][]{Boogert00b}.
If our understanding of the evolution of a protostar is basically
correct, those protostars, typically Class I or border line Class II
sources, represent a relatively evolved stage, where most of the
original envelope has already been dispersed \citep[e.g.][]{Shu87,
  Andre00}.  Furthermore, the observed absorption may be dominated by
foreground molecular clouds \citep{Boogert02}.  Thus, direct
observations of the chemical composition of the primeval dust mantles
of low mass protostars have so far proven to be elusive.
 
Alternatively, one can carry out an ``archeological'' study, looking
at the composition of the gas in the regions, which are known or
suspected to be dominated by the gas desorbed from grain mantles.
This technique has the advantage of being much more sensitive than the
absorption technique, as it can detect molecules whose abundance (with
respect to H$_2$) is as low as $\sim 10^{-11}$ against a limit of
$\sim 10^{-6}-10^{-7}$ reachable with the absorption technique.
Indeed, several very complex molecules observed in the warm ($\geq
100$ K) gas of the so called \emph{hot cores} have been considered
hallmarks of grain mantle evaporation products
\citep[e.g.][]{Blake87}.  Once in the gas phase, molecules like
formaldehyde and methanol, initially in the grain mantles, trigger the
formation of more complex molecules, referred to as daughter or
second-generation molecules \citep[e.g.][]{Charnley92, Caselli93}.
The gas temperature and density are other key parameters in the
chemical evolution of the gas, which has the imprint of the
pre-collapse phase.
 

So far, hot cores have been observed in massive protostars, and are
believed to represent the earliest stages of massive star formation,
when the gas is not yet ionized by the new born star \citep{Kurtz00}.
Recently, however, it has been proposed that low mass protostars might
also harbor such hot cores. Note that the definition of hot core is
not unanimous in the literature. Here we mean a region where the
chemical composition reflects the evaporation of the ice mantles and
subsequent reactions between those species \citep[e.g.][]{Rodgers03}.
In this respect, \citet{Ceccarelli00a,Ceccarelli00b} claimed that the
low mass protostar \object{IRAS16293-2422} shows evidence of an inner
region ($\sim 400$ AU in size) warm enough ($\geq$ 100 K) to evaporate
the grain mantles, a claim substantially confirmed by
\citet{Schoier02}.  Indeed, very recent observations by Cazaux et al.
(\citeyear{Cazaux03}; see also \citenp{Ceccarelli00c}) reveal also the
presence of organic acids and nitriles in the core of
\object{IRAS16293-2422}, substantiating the thesis of a hot core
region in which not only the ices have evaporated but also a
subsequent hot core chemistry has ensued.  Furthermore,
\citet{Maret02} argued that \object{NGC1333-IRAS4}, another low mass
very embedded protostar, has also such a warm region, somewhat less
than 200 AU in size.
 
Formaldehyde is a relatively abundant constituent of the grain mantles
and it is a basic organic molecule that forms more complex molecules
\citep[e.g.][]{Charnley92}.  For this reason, we studied the
formaldehyde line emission originating in the envelopes of a sample of
very embedded, Class 0 low mass protostars.  In this article we report
the first results of this systematic study.  This is part of a larger
project aimed to characterize as far as possible the physical and
chemical composition of low mass protostars during the first phases of
formation. \citet{Jorgensen02} determined the temperature and density
structure for these sources and the CO abundance in the outer regions.
A forthcoming paper will address the methanol line emission in the
same source sample, as methanol is another key organic mantle
constituent, linked by a common formation route with formaldehyde.
 
One of the ultimate goals of the present study is to understand the
efficiency of H$_2$CO against CH$_3$OH formation in low mass
protostars, whether and how it depends on the parental cloud, and to
compare it with the case of massive protostars.  An immediate goal of
the present article is to study the formaldehyde abundance profile in
the surveyed sample of low mass protostars.  In a previous study that
we carried out towards \object{IRAS16293-2422}, we concluded that
formaldehyde forms on grain mantles and is trapped mostly in
H$_2$O-rich ices in the innermost regions of the envelope and mostly
in CO-rich ices in the outermost regions
\citep{Ceccarelli00b,Ceccarelli01}.  As the dust gradually warms up
going inwards, formaldehyde is released from the icy mantles all along
the envelope.  In the hot core like region ($r \leq 200$ AU) the
formaldehyde abundance jumps by about a factor 100 to $\sim 1 \times
10^{-7}$ \citep{Ceccarelli00b,Schoier02}. Similarly, formaldehyde
enhancement is observed in several outflows, because of ice mantle
sputtering in shocks \citep{Bachiller97, Tafalla00}. In contrast, no
jump of formaldehyde abundance has been detected in the sample of
massive protostars studied by \citet{vanderTak00}. To firmly assess
whether and by how much formaldehyde is systematically more abundant
in the interiors of low compared to high mass protostars, a survey of
more low mass protostars has to be carried out.  This will allow us to
answer some basic questions such as how, when and how much
formaldehyde is formed on the grain mantles. Given that it forms more
complex molecules \citep[e.g.][]{Bernstein99} knowing the exact
abundance of formaldehyde is fundamental to answer the question of
whether or not pre- and/or biotic molecules can be formed in the 200
or so inner AUs close to the forming star.
 
In this article we report observations of formaldehyde emission in a
sample of eight Class 0 sources.  After a preliminary analysis
(rotational diagrams and LVG analysis), the observations are analyzed
in terms of an accurate model that accounts for the temperature and
density gradients in each source, as well as the radiative transfer,
which includes FIR photon pumping of the formaldehyde levels.  The
article is organized as follows: we first explain the criteria that
lead to the source and line selection and the observations carried out
(\S 2).  In \S 3 we describe the results of the observations, in \S 4
we derive the approximate gas temperature, density and H$_2$CO column
density of each source by means of the standard rotational diagram
technique and by a non-LTE LVG model. In \S 5 we derive the
formaldehyde abundance in the inner and outer parts of the envelope of
each source, with an accurate model that takes into account the
structure of the protostellar envelopes.  Finally, in \S 6 we discuss
the implications of our findings, and conclude in \S 7.
 
\section{Observations} 
\label{observations} 
 
\subsection{Target and line selection} 
 
We observed a sample of eight protostars, all of them \emph{Class 0}
sources \citep{Andre00} located in the Perseus, $\rho$-Ophiuchus and
Taurus complexes, except \object{L1157-MM} that lies in an isolated
clump \citep{Bachiller97}.  The eight selected sources are among the
brightest Class 0 sources in the \citet{Andre00} sample.  Their
physical structure (dust density and temperature profiles) has been
determined from their continuum emission by \citet{Jorgensen02}
except for \object{L1448-N}, which is analyzed in this paper
(see Appendix \ref{sec:dens-temp-prof}).
 
The source distances quoted by \citet{Jorgensen02} were adopted.  The
list of the selected sources is reported in Table \ref{sources}
together with their bolometric luminosity, envelope mass, the ratio of
the submillimeter to bolometric luminosity, and the bolometric
temperature and distance.  In the same Table, we also report the data
relative to \object{IRAS16293-2422}, which was previously observed in
H$_2$CO transitions by \citet{vanDishoeck95} and \citet{Loinard00},
and studied in \citet{Ceccarelli00b} and \citet{Schoier02}.
\object{IRAS16293-2422} will be compared to the other sources of the
sample.
 
\input{table_sources}

A list of eight transitions was selected, three ortho-H$_2$CO
transitions and five para-H$_2$CO transitions (Table
\ref{table_fluxes_h2co}). When possible, the corresponding isotopic
lines were observed in order to derive the line opacity.  The
transitions were selected to cover a large range of upper level
energies (between $\sim$ 20 and $\sim$ 250 K) with relatively large
spontaneous emission coefficients $A_{u,l}$ ($\geq 10^{-4}$ s$^{-1}$).
The latter condition is dictated by the necessity to detect the line,
whereas the first condition aims to obtain lines that probe different
regions of the envelope.  Finally, practical considerations, namely
having more than one line in a single detector setting, provided a
further constraint.  In the final selection we were helped by our
pilot study on \object{IRAS16293-2422} \citep{Ceccarelli00b} and by
previous studies of the formaldehyde emission in molecular clouds and
protostellar envelopes
\citep{Mangum93,Jansen94,Jansen95,Ceccarelli03}.
 
The formaldehyde transitions between 140 and 280 GHz were observed
with the single dish IRAM-30m telescope\footnote{IRAM is an
  international venture supported by INSU/CNRS (France), MPG (Germany)
  and IGN (Spain).}, located at the summit of Pico Veleta in Spain.
Higher frequency lines were observed at the JCMT\footnote{The JCMT is
  operated by the Joint Astronomy Center in Hilo, Hawaii on behalf of
  the present organizations: The Particle Physics and Astronomy
  Research Council in the United Kingdom, the National Research
  Council of Canada and the Netherlands Organization for Scientific
  Research.}, a 15 meter single dish telescope located at the summit
of Mauna Kea, Hawaii.  The choice of the two telescopes allows us to
have roughly similar beam sizes over the observed frequencies.
  
\subsection{IRAM observations} 

The IRAM observations were carried out in November 1999\footnote{IRAM
November 1999 data have also been presented in \citet{Loinard02a}} and
September 2002. The various receivers available at the 30-m were used
in different combinations to observe at least four transitions
simultaneously. The image sideband rejection was always higher than 10
dB, and typical system temperatures were 200-300 K at 2 mm, and
200-500 K at 1 mm. The intensities reported in this paper are
expressed in main beam temperature units, given by:
 
 
\begin{equation} 
T_\mathrm{mb} = \frac{F_\mathrm{eff}}{B_\mathrm{eff}} T_\mathrm{A}^{*} 
\end{equation} 
 
\noindent
where $B_\mathrm{eff}$ is the main beam efficiency, and $F_{\rm eff}$
is the forward efficiency. The main beam efficiency is 69\%, 57\% and
42\% at 140, 220 and 280 GHz respectively, and the forward efficiency
is 93\%, 91\% and 88\% at the same frequencies. Each receiver was
connected to an autocorrelator unit. For the 1 and 2 mm bands, a
spectral resolution of 80 kHz and a bandwidth of 80 MHz was used.
These spectral resolutions correspond to a velocity resolution of
0.09-0.17 km s$^{-1}$ depending on frequency.  All IRAM observations
were obtained in position switching mode. The absolute calibration was
regularly checked and was about 10\%, 15\% and 20\% at 140-170 GHz,
220-240 GHz and 280 GHz respectively. Pointing was also regularly
checked and was better than 3''.
 
\subsection{JCMT observations} 
 
The JCMT observations were obtained in February 2001, August 2001 and
February 2002. The single sideband dual polarization receiver B3 was
used with the Digital Autocorrelation Spectrometer (DAS). Typical
system temperatures were 400 to 800 K. A spectral resolution of 95 kHz
for a 125 MHz bandwidth was used for most of the lines, while a
resolution of 378 kHz for a bandwidth of 500 MHz was used to observe
some of the lines simultaneously. These spectral resolutions
correspond to a velocity resolution of 0.08-0.32 km s$^{-1}$. The
antenna temperatures were converted into main beam temperature scale
using\footnote{JCMT does not follow the same convention for antenna
  temperature than IRAM, the JCMT antenna temperature
  $T_\mathrm{A}^{*}$ being already corrected for the forward
  efficiency of the antenna.}
 
\begin{equation} 
T_\mathrm{mb} = \frac{T_\mathrm{A}^{*}}{\eta_\mathrm{mb}} 
\end{equation} 

\noindent
where $\eta_\mathrm{mb}$ is the main beam efficiency, equal to 63\% at
the observed frequencies. The calibration and pointing were regularly
checked using planets and were found to be better than 30\% and
3'' respectively. The JCMT observations were obtained in beam
switching mode, with a 180'' offset.
  
\section{Results} 
 
\begin{figure*} 
  \centering{ 
    \includegraphics[height=20cm]{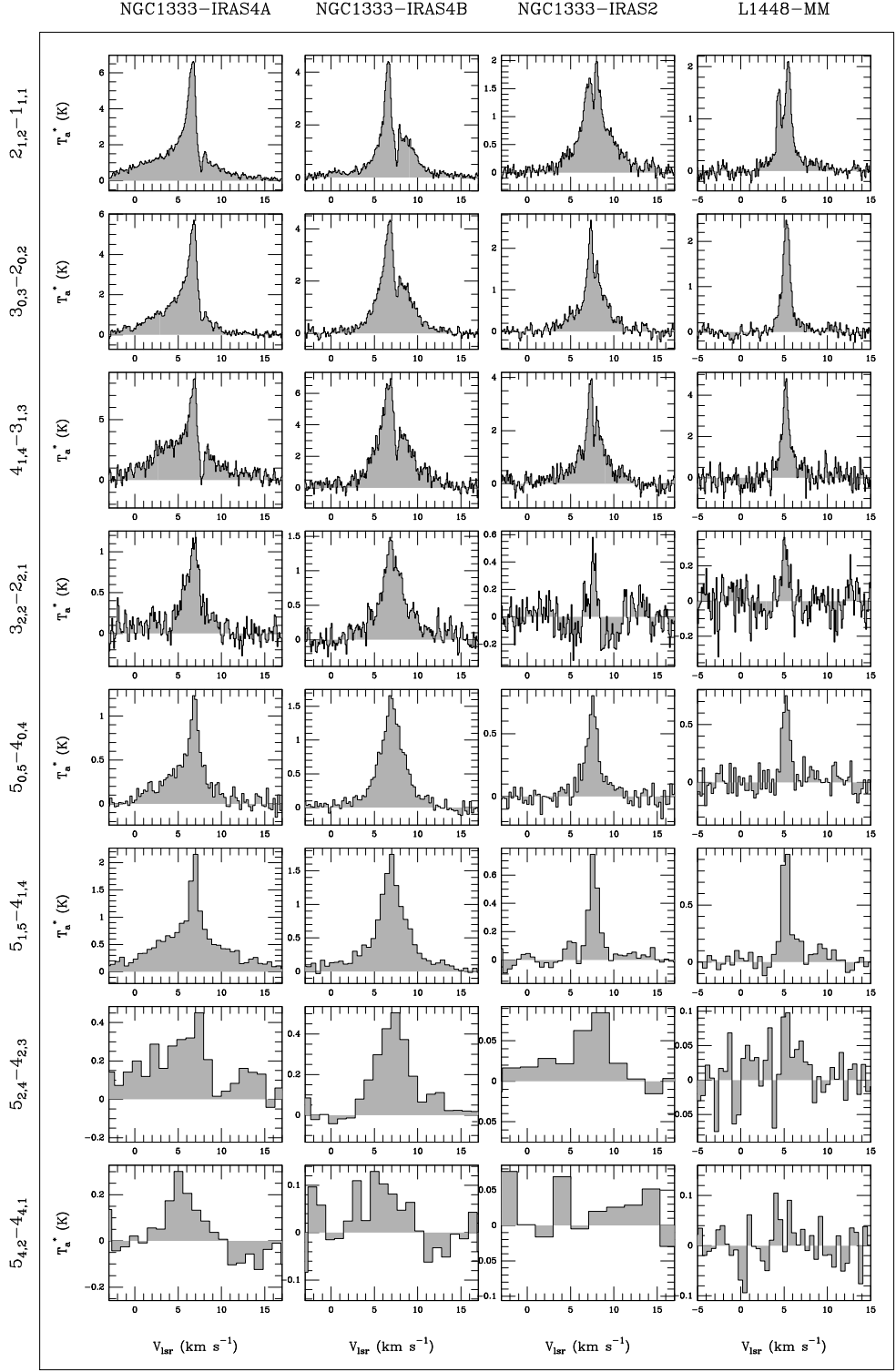} 
  } 
  \caption{Spectra of the eigth H$_2$CO transitions in Table 
    \ref{table_fluxes_h2co} observed towards \object{NGC1333-IRAS4A}, 
    \object{NGC1333-IRAS4B}, \object{NGC1333-IRAS2}, and 
    \object{L1448-MM} respectively.} 
  \label{fig_spectra_h2co_1}  
\end{figure*}

\begin{figure*} 
  \centering{ 
    \includegraphics[height=20cm]{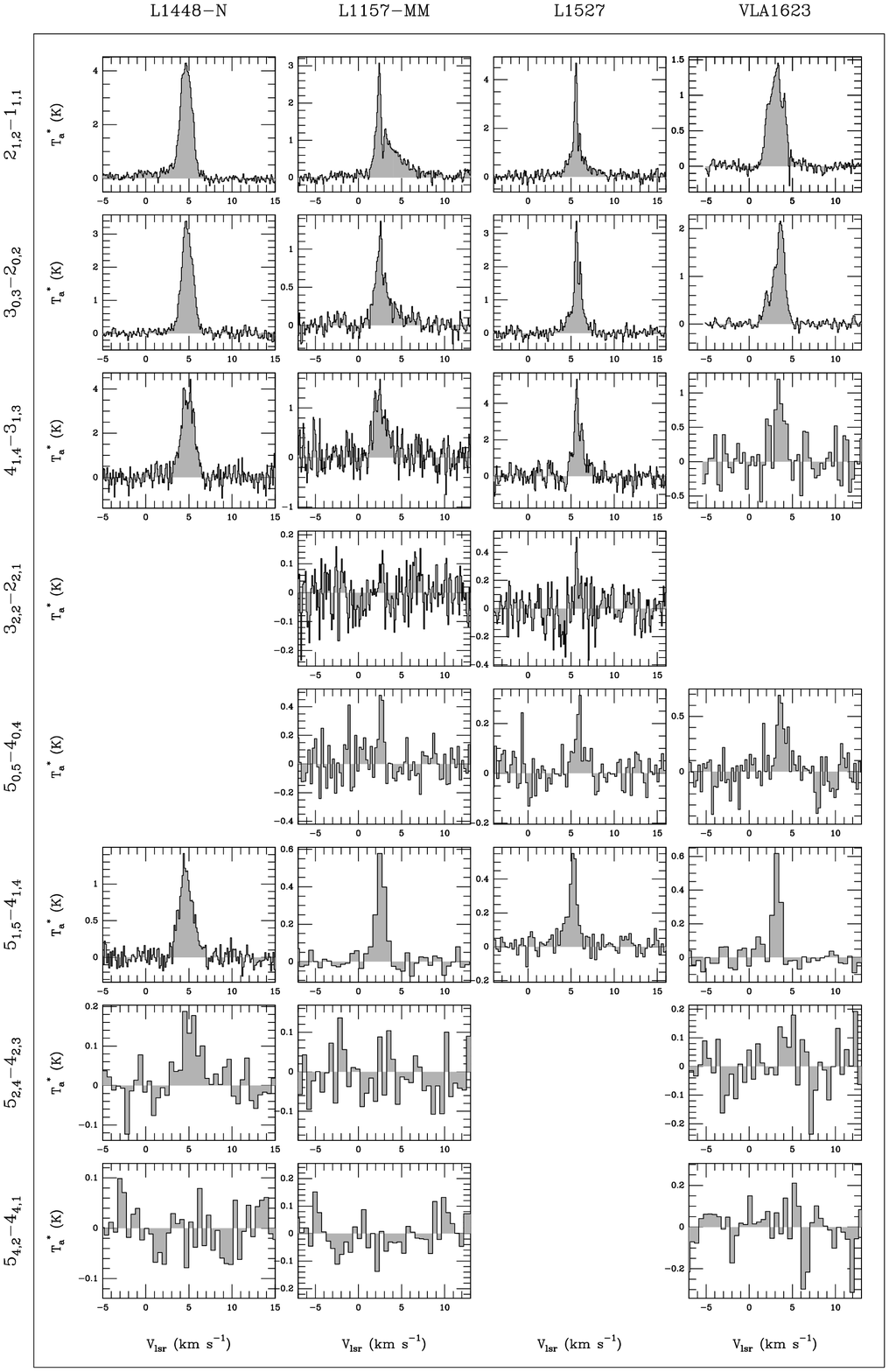} 
  }  
    \caption{As Fig. \ref{fig_spectra_h2co_2} for the sources
      \object{L1448-N}, \object{L1157-MM}, \object{L1527} and
      \object{VLA1623}.} 
  \label{fig_spectra_h2co_2}  
\end{figure*}

The observed H$_2$CO line spectra are shown in
Figs. \ref{fig_spectra_h2co_1} and \ref{fig_spectra_h2co_2} and the
results of the observations are summarized in Table
\ref{table_fluxes_h2co}.
 
Most of the lines are relatively narrow (FWHM $\sim$ 2-3 km s$^{-1}$)
with a small contribution ($\leq$ 5\%) from wings extending to larger
velocities.  The higher the upper level energy of the transition the
lower the contribution of the wings, which practically disappear in
the lines observed with JCMT. \object{NGC1333-IRAS4A} and
\object{NGC1333-IRAS4B} represent an exception to this picture.  The
line spectra of these two sources are broad ($\sim$ 5 km s$^{-1}$) and
the wings are more pronounced than in the other sources.  Evidence of
self-absorption and/or absorption from foreground material is seen in
most sources, in particular in low lying lines.

\input{table_flux_h2co}

In this study we focus on the emission from the envelopes surrounding
the protostars.  Hence, we are interested in the intensity of the
narrow component of the lines, that we fitted with a Gaussian.  In
some cases, a residual due to the ``high'' velocity wings remains, and
that has not been included in the line flux estimate. When a Gaussian
fitting was not possible because of self-absorption, the flux of the
lines was estimated by integrating over a velocity range of $\pm$ 2 km
s$^{-1}$ around the source $V_\mathrm{lsr}$. This velocity range
corresponds to the width of the lines with high upper level energies,
where self-absorption is less important. For these lines, the
self-absorption is included in the line flux determination, and the
flux measured is therefore slightly smaller than the one that would
have been obtained by a Gaussian fitting. We observed the brightest
lines in the $^{13}$C isotopomer of formaldehyde, as reported in Table
\ref{table_fluxes_h213co}.
 
\begin{figure*}  
    \centering{ 
      \includegraphics[height=17cm,angle=270]{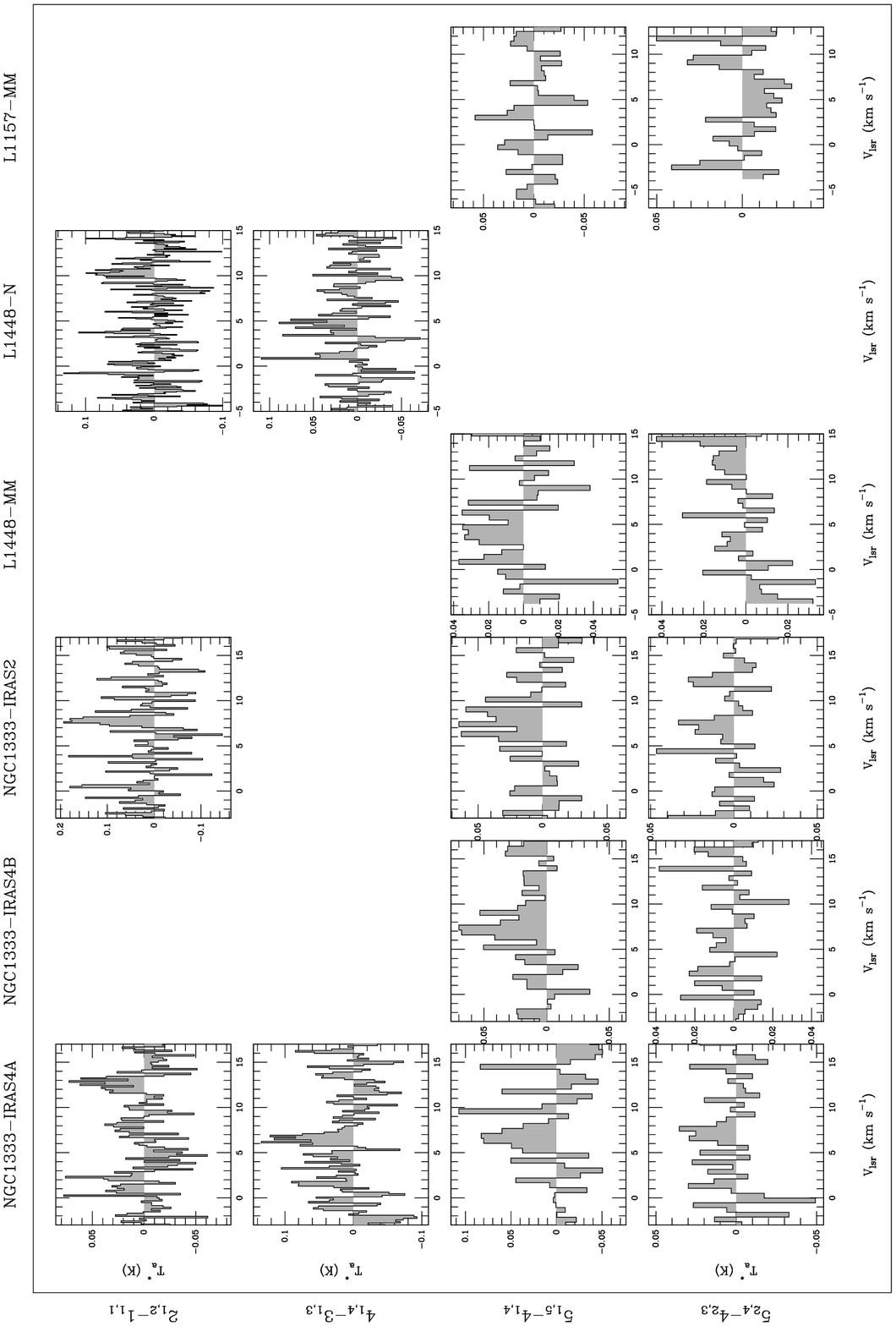} 
    }  
    \caption{Spectra of the observed H$_2^{13}$CO lines.} 
    \label{fig_spectra_h213co}  
\end{figure*} 
 
Finally, the errors quoted in Table \ref{table_fluxes_h2co} and
\ref{table_fluxes_h213co} include both the statistical uncertainties
and the calibration error. For non detected lines we give the
2$\sigma$ upper limit defined as follows:
 
\begin{equation} 
  F_\mathrm{max} = 2 (1 + \alpha) \, \sigma \, \sqrt{\Delta v \,
  \delta v} 
\end{equation} 

\noindent
where $\sigma$ is the RMS per channel, $\Delta v$ is the line width
estimated from detected lines on the same source, $\delta v$ is the
channel width, and $\alpha$ is the calibration uncertainty.
 
\input{table_flux_h213co}

\section{Approximate analysis} 
 
\subsection{Line opacities} 
 
The detection of some H$_2^{13}$CO transitions allows to estimate the 
line opacities of the relevant H$_2^{12}$CO lines. Using the escape 
probability formalism and assuming that the H$_2^{13}$CO lines are 
optically thin, the ratio between the H$_2^{13}$CO and H$_2^{12}$CO 
line fluxes can be expressed as: 
 
\begin{equation} 
  \frac{F_\mathrm{H_{2}^{12}CO}}{F_\mathrm{H_{2}^{13}CO}} = 
  \frac{[^{12}\mathrm{C}]}{[^{13}\mathrm{C}]} \beta 
\end{equation} 

\noindent
where $\beta$ is the escape probability, which, in the case of a
homogeneous slab of gas \citep{deJong80}, is equal to:
 
\begin{equation}  
  \beta = \frac{1-\exp(-3\tau)}{3\tau} 
\end{equation} 
 
\noindent
and $\frac{[^{12}\mathrm{C}]}{[^{13}\mathrm{C}]}$ is the isotopic
elemental ratio, equal to 70 \citep{Boogert00a}. In the previous
equation, we assume that the H$_2^{12}$CO to H$_2^{13}$CO ratio is
equal to the isotopic elemental ratio, as supported by the available
observations \citep[e.g.][]{Schoier02}. Using this equation, the
opacities values reported in Table \ref{table_opacities} are obtained.
 
\input{table_opacities}

The opacity values reported in Table \ref{table_opacities} are
relatively low, which indicates that the lines are moderately thick,
vwith the exception of the line at 351 GHz towards
\object{NGC1333-IRAS2}. The uncertainty on the latter opacity is however
relatively large, as shown by the errors bars reported in Table
\ref{table_opacities}.
 
\subsection{Rotational diagram analysis} 
\label{popdiag_analysis} 
 
To obtain a first order estimate, we derived the beam-averaged column
density of formaldehyde and rotational temperature by means of the
standard rotational diagram technique (see \citenp{Goldsmith99} for a
general description of the method, and \citenp{Mangum93} for its
application to formaldehyde lines).
 
 
%
%
%
%
%
%
%
%
 
\begin{figure*} 
  \centering{ 
    \includegraphics[width=17cm]{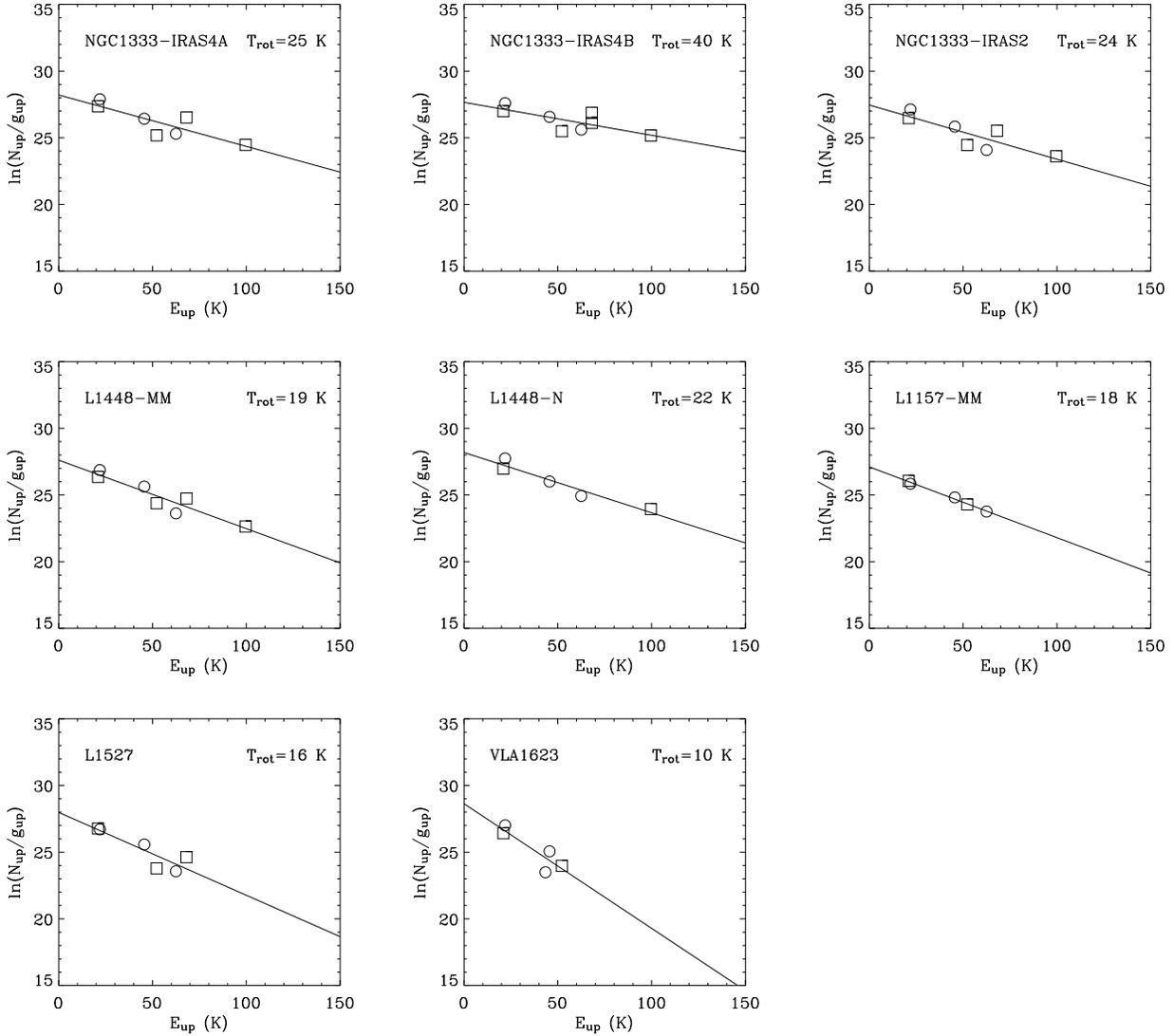} 
  } 
    \caption{H$_2$CO rotational diagrams derived for the observed 
      sources. Circles and squares mark the ortho and para H$_2$CO 
      transitions respectively. Fluxes of the para transitions have been 
      multiplied by the ortho to para ratio, obtained by minimizing 
      the $\chi^2$ between the observations and the predictions of the 
      rotational diagram (see text). Solid lines show the best fit 
      curves.} 

 
  \label{fig_popdiag}  
\end{figure*}

Fig. \ref{fig_popdiag} shows the H$_2$CO rotational diagrams of the
observed sources.  In these diagrams, the ortho to para ratio of
formaldehyde was kept as a free parameter, and was derived by
minimizing the $\chi^2$ between the observed fluxes and the rotational
diagram predictions. The best agreement is obtained for a value of
about 2 on all the sources. The fact that this value is lower than the
high-temperature limit of 3 suggest that the formaldehyde is formed at
low temperature, around 20 K \citep{Kahane84}. However, while this
ratio seems to be the same for all sources, we emphasize that it is
highly uncertain.  In particular, the fact that the same transitions
have been observed on all the sources can lead to systematic errors on
this value. A more accurate derivation of the ortho to para ratio
would need a correction for the line opacities, which has the effect
of scattering the points in the rotational diagram \citep[see][ for a
review on the effect of opacities in a rotational
diagram]{Goldsmith99}. This correction is not possible here as only a
limited number of H$_2^{13}$CO transitions has been observed. The
5$_{2,4}$-4$_{2,3}$ line was only detected towards
\object{NGC1333-IRAS4A} and \object{NGC1333-IRAS4B}. Because of its
high energy (234 K), this line is probably excited in hotter regions,
and would increase the derived rotational temperature. This line was
not included to keep the derived parameters comparable from one source
to the other.
 
Table \ref{table_results_LTE_LVG} summarizes the derived total column
densities and rotational temperatures. The column densities range from
2 to $7 \times 10^{13}$ cm$^{-2}$, and the rotational temperatures
from 11 to 40 K. The values are both only lower limits to the actual
gas temperature and column density, as the gas temperatures can
actually be significantly higher in the case of non-LTE conditions,
and the derived column density can also be higher in case of optically
thick emission. To correct for this effect, the derived column
densities were recalculated adopting the average value of opacities
quoted in Table \ref{table_opacities}.  The corrected column densities
are also reported in Table \ref{table_results_LTE_LVG}, and range from
0.8 to 2 $\times 10^{14}$ cm$^{-2}$.

\input{table_result_LTE_LVG}

\subsection{LVG modeling} 
 
In order to derive the physical conditions of the emitting gas under
non-LTE conditions, the formaldehyde emission has been modeled using
an LVG code\footnote{Details on the used LVG code can be found in
\citet{Ceccarelli02}.}. The collisional coefficients from
\citet{Green91} and the Einstein coefficients from the JPL database
\citep{Pickett98} were used. The LVG code has three free parameters:
the column density to line width ratio
$\frac{N(\mathrm{H_{2}CO})}{\Delta v}$ (which regulates the line
opacity), the gas temperature $T_\mathrm{gas}$, and the molecular
hydrogen density $n(\mathrm{H_2})$. When the lines are optically thin
the line ratios only depend on the latter two parameters.  Since we
measured only marginally optically thick lines, the gas temperature
and density were first constrained based on the line ratios predicted
in the case of optically thin lines. The absolute line fluxes
predicted by the model were then compared with observations to
constrain the H$_2$CO column density.
 
\begin{figure*}  
    \centering{
      \includegraphics[width=17cm]{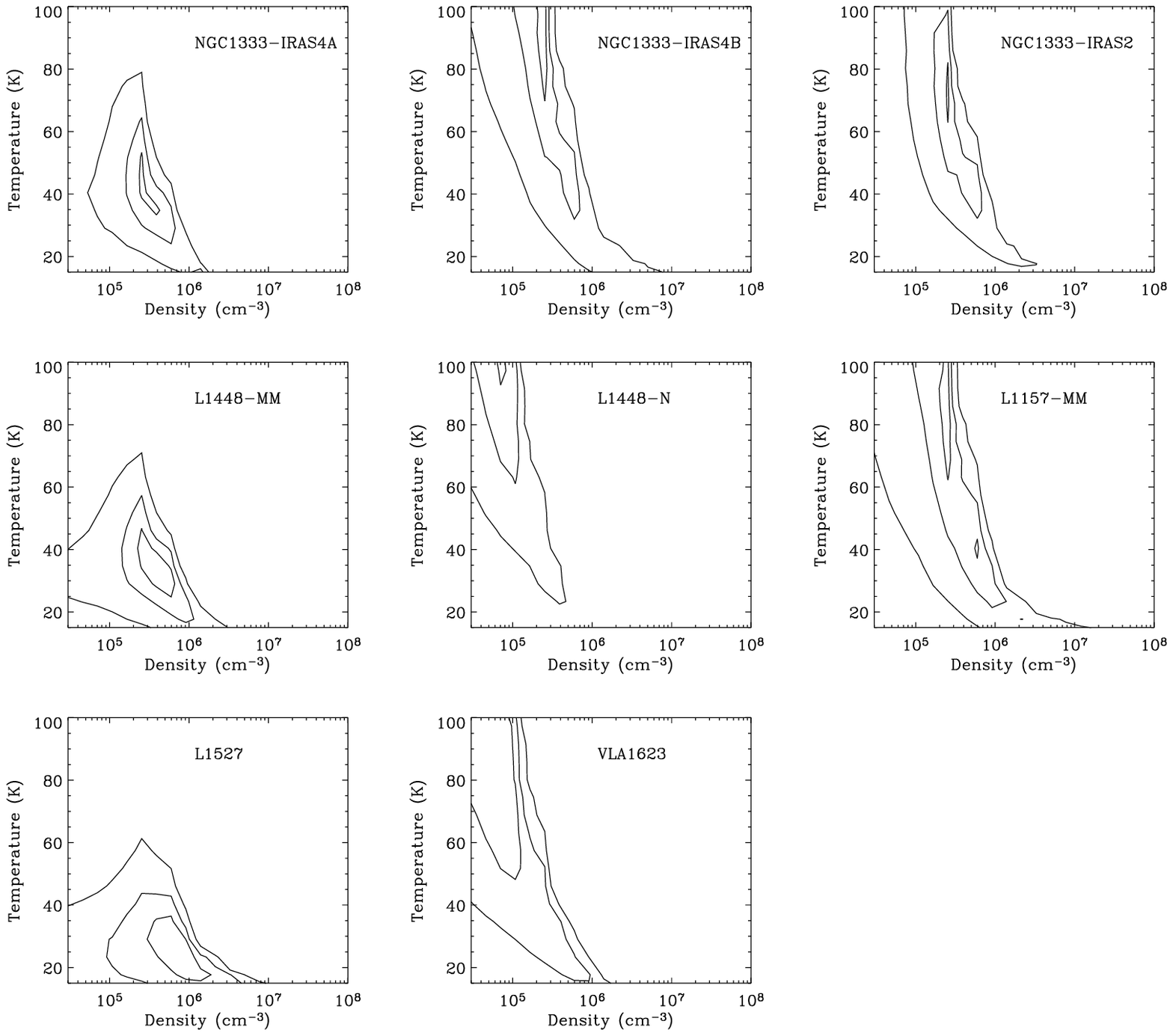}  
      \caption{$\chi^{2}$ contours as a function of the density and
      temperature of the emitting gas. Contours indicates the 1, 2 and
      3$\sigma$ confidence levels respectively.}
      \label{fig_chi2_dens_temp}
    }  
\end{figure*}

The gas temperature and density have been determined by minimizing the
$\chi_{\rm red}^2$, defined as:
 
\begin{equation} 
\chi_{\rm red}^2 = \frac{1}{N-2} \sum_1^N \frac{\left(
  \mathrm{Observations - Model} \right)^2}{\sigma^2} 
\end{equation}

\noindent 
where all the observed H$_2$CO lines were
included. Fig. \ref{fig_chi2_dens_temp} show the $\chi^2$
contours. The derived $T_\mathrm{gas}$ and $n(\mathrm{H_2})$ are
reported in Table \ref{table_results_LTE_LVG}. The H$_2$CO column
densities were then constrained using the observed o-H$_2$CO
5$_{1,5}$-4$_{1,4}$ line flux, under the assumption of optically thin
emission.

The derived temperatures are between 30 and 90 K. These temperatures
are significantly higher than the rotational temperature, suggesting
that the observed transitions are subthermally populated.  Indeed, the
derived densities support this conclusion, as they vary between 1 and
6 $\times$ 10$^{5}$ cm$^{-3}$.  It is remarkable that the density
range is so small, but the density estimate is certainly biased
because of the choice of the transitions. In particular, the densities
are slightly lower than those found by \citet{Blake95} and
\citet{vanDishoeck95} for \object{NGC1333-IRAS4A},
\object{NGC1333-IRAS4B}, and \object{IRAS16293-2422}. This is probably
due to the fact that those studies included only the higher frequency
lines whereas we here included also lower frequency, and thus lower
critical density lines. This re-inforces the conclusion that a range
of densities are present in the envelope, as predicted by the
power-law density structure derived from continuum observations
\citep{Jorgensen02}. Finally, the H$_2$CO column densities derived
with the rotational diagram method are typically lower by less than a
factor 5 than the ones derived using the LVG method with the exception of
\object{NGC1333-IRAS2}.
 
Table \ref{table_results_LTE_LVG} also reports estimates of the
average H$_2$CO abundance in each source, obtained dividing the
H$_2$CO by the H$_2$ column densities derived by \citet{Jorgensen02}
from submillimeter continuum observations.  The latter refer to the
amount of material with a temperature larger than 10 K, typically at a
distance of several thousands of AUs, more than the envelope
encompassed by the beam of our observations. This material is likely
an upper limit to the column density of the gas emitting the H$_2$CO
lines, but it provides a first approximate estimate of the H$_2$CO
abundance.  Table \ref{table_results_LTE_LVG} shows rather large
variations in the H$_2$CO abundance from source to source.  Given the
approximation of the method used to derive them, this spread may not
be entirely real.  In the next section we analyze the observed lines
by means of an accurate model that takes into account the source
structure.
 
 \section{Protostellar envelope model} 
\label{envelope_model} 
 
\subsection{Model description} 
 
The model used computes the line emission from a spherical envelope.
Dust and gas have density and temperature gradients, that are
approximated as follows. The density profile is described by a power
law $n(\mathrm{H_2})\propto r^{-\alpha}$, where $\alpha$ is between
0.5 and 2. The case $\alpha = 1.5$ corresponds to an entirely
free-falling envelope, whereas $\alpha = 2$ would mimic an isothermal
sphere in hydrostatic equilibrium. The densities and dust temperature
profiles of all the sources of the sample have been derived by
\citet{Jorgensen02}, except for \object{L1448-N}, whose analysis is
reported in Appendix \ref{sec:dens-temp-prof}.

\begin{figure}  
    \centering{
      \includegraphics[width=\hsize]{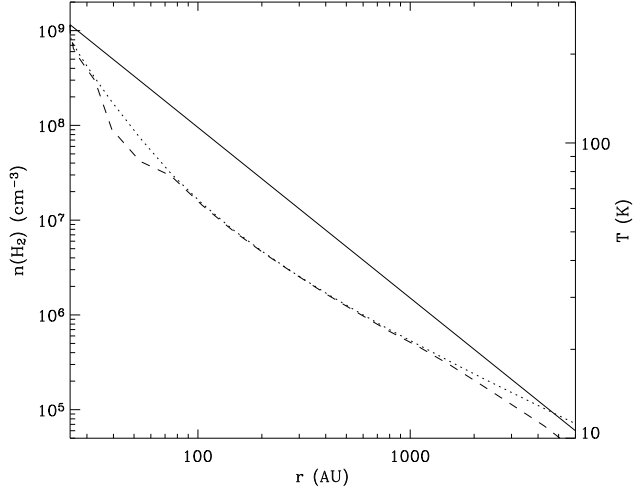}  
      \caption{Density (solid line), gas (dashed line) and dust (dotted 
        line) temperature across the envelope of \object{NGC1333-IRAS2}.} 
      \label{fig_denstemp_iras2}
    }  
\end{figure}

The gas temperature profile has then been computed by using the model
developed by \citet{Ceccarelli96}, which solves the thermal balance in
the envelope. In order to compute the gas temperature, one needs to
solve the radiative transfer of the main coolants of the gas, i.e.
H$_{2}$O, CO and O. For the water abundance in the inner and outer
regions we used the values derived by the analysis of
\object{IRAS16293-2422} \citep{Ceccarelli00a} and
\object{NGC1333-IRAS4} \citep{Maret02}: 4 $\times$ 10$^{-6}$ and 4
$\times$ 10$^{-7}$ respectively. The CO abundance in the outer region
is taken to be 10$^{-5}$ \citep[e.g.][]{Jorgensen02}, lower than the
canonical value, as this species is depleted in the cold parts of the
envelope. Finally, the oxygen abundance is taken to be 2.5 $\times$
10$^{-4}$. Fig. \ref{fig_denstemp_iras2} shows as a typical example
the case of \object{NGC1333-IRAS2}. The gas temperature tracks
closely, but not completely, the dust temperature. In the very inner
region the gas is colder than the dust because of the increase of the
water abundance caused by the icy mantle evaporation (when
$T_\mathrm{dust}$ $\geq$ 100 K), which increases the gas cooling
capacity.  In the very outer region the gas is colder than the dust
too, because of the efficient gas cooling by the CO lines (see also
the discussion in \citenp{Ceccarelli96} and \citenp{Maret02}). These
differences concern however small regions in the envelope, and
therefore the results would be essentially the same if the gas and
dust temperature are assumed to be equal. To fully quantify this
effect, we ran a model for \object{NGC1333-IRAS2} with a gas
temperature equal to the dust one, and no significant differences were
found.
 
The velocity field, which regulates the line opacity in the inner
envelope, has been approximated as due to free falling gas towards a
0.5 M$_\odot$ central object in all sources (no turbulent broadening
is taken into account). In view of the importance of ice evaporation,
the formaldehyde abundance across the envelope has been modeled by a
step function: when the dust temperature is lower than the ice mantle
evaporation ($T_\mathrm{dust}$ $\leq$ 100 K) the abundance has the
value \Xout, whereas it increases to \Xin in the $T_\mathrm{dust}$
$\geq$ 100 K region.  Finally, a H$_2$CO ortho to para ratio of 3 was
assumed\footnote{Since the value of 2 obtained with the rotational
diagrams is highly uncertain, we prefer to adopt the canonical value
of 3 for the standard model. We will show in section
\ref{sec:h_2co-ortho-para} that the results are not substantially
affected by this value.}. We will discuss the dependence of the
obtained results on these ``hidden'' parameters in the next section.
 
Finally, the line emission is computed by solving the radiative
transfer in presence of warm dust in the escape probability formalism
where the escape probability $\beta$ is computed at each point of the
envelope by integrating the following function over the solid angle
$\Omega$ \citep{Ceccarelli96}:
 
\begin{equation} 
  \beta = \frac{k_\mathrm{d}}{k_\mathrm{L} + k_\mathrm{d}} + 
  \frac{k_\mathrm{L}}{(k_\mathrm{L} + k_\mathrm{d})^2} \int d\mu 
  \frac{1-\exp \left[ - \left( k_\mathrm{L} + k_\mathrm{d} \right) 
  \Delta L_\mathrm{th} \right]} {\Delta L_\mathrm{th}} 
\end{equation} 

\noindent  
where $k_\mathrm{L}$ and $k_\mathrm{d}$ are the line and dust
absorption coefficients respectively, and $\Delta L_\mathrm{th}$ is
the line trapping region, given by the following expressions:
 
\begin{equation} 
  \Delta L_\mathrm{th} = 2 \Delta v_\mathrm{th}  
  \left( \frac{v}{r} \left| 1-\frac{3}{2} \mu^2 \right| \right)^{-1}  
\end{equation} 


\noindent
in the infalling region of the envelope (where $\mathrm{arcos} \left(
\mu \right)$ is the angle with the radial outward direction) and
 
\begin{equation} 
  \Delta L_\mathrm{th} = r \left( 1 - \frac{r}{R_\mathrm{env}} \right) 
\end{equation} 
 
\noindent 
in the static region (where $R_\mathrm{env}$ is the envelope
radius). In the previous equations, $\Delta v_\mathrm{th}$ is the
thermal velocity width and $v$ is the infall velocity. In practice,
the photons emitted by the dust can be absorbed by the gas and can
pump the levels of the formaldehyde molecules.  At the same time,
photons emitted by the gas can be absorbed by the dust (at the
submillimeter wavelengths the dust absorption is however negligible).
  
\subsection{Results} 

\begin{figure*}  
  \centering{  
    \includegraphics[width=17cm]{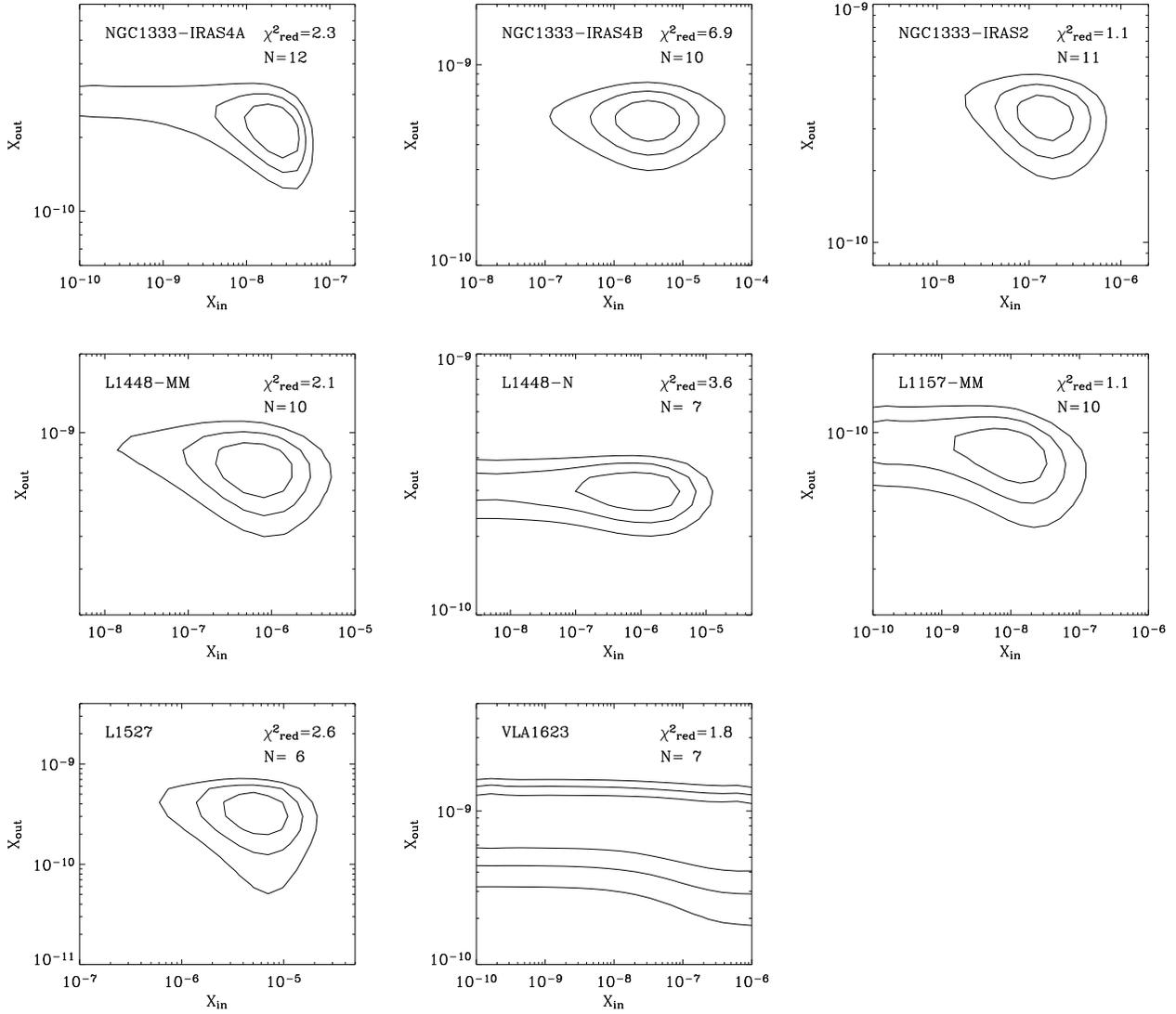} 
    \caption{Protostellar envelope model $\chi_{\rm red}^2$ contours as a 
      function of the outer and inner H$_2$CO abundances. The contours 
      levels show the 1, 2, and 3 $\sigma$ confidence levels 
      respectively.}
    \label{fig_chi2_xin_xout}  
  }
\end{figure*}

In order to constrain the inner and outer formaldehyde abundance in
the envelope, we run a grid of models, varying \Xout between
$10^{-12}$ and $10^{-8}$, and \Xin between $10^{-12}$ and $10^{-4}$
respectively for each source. Both parameters were constrained by
minimizing the $\chi_{\rm red}^2$. The best fit model for each source
was then obtained from the $\chi_{\rm red}^2$ diagrams shown in Fig.
\ref{fig_chi2_xin_xout}, and the parameters are summarized in Table
\ref{table_result_infall}.  The list of predicted o-H$_2$CO spectra
for each source are reported in Appendix \ref{sec:models-pred-form}.

\input{table_result_infall} 
 
\Xout is well constrained in all sources, and varies between $8 \times
10^{-11}$ and $8 \times 10^{-10}$.  These values differ on average by
a factor 3 from the abundances derived by the LVG analysis.  In four
sources (NGC1333-IRAS4B, NGC1333-IRAS2, L1448-MM and L1527) the value
of \Xin is also well constrained by the observations, with a 3$\sigma$
confidence level.  In three sources (NGC1333-IRAS4A, L1448-N and
L1157-MM) we also detected formaldehyde abundance jumps, but the level
of confidence is less or equal to 2$\sigma$.  VLA1623 is the only
source where no evidence of a jump is detected, although the data
would not be inconsistent with it.  The measured \Xin values are
between $1 \times 10^{-8}$ and $6 \times 10^{-6}$, and the jumps in
the formaldehyde abundance are between 100 and $10^4$.

\begin{figure}  
  \centering{  
    \includegraphics[width=\hsize]{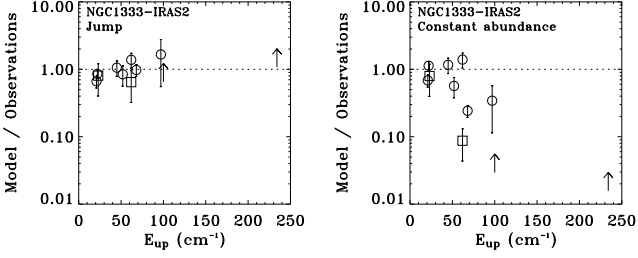} 
  } 
  \caption{Ratio of the model predictions over the observed fluxes
    of H$_2$CO lines as a function of the upper energy of the line,
    for \object{NGC1333-IRAS2}. In the left panel a jump in the
    abundance (Tab. \ref{table_result_infall}) is adopted, while the
    in the right panel a constant abundance across the envelope is
    assumed.  The circles and the squares represent H$_2^{12}$CO and
    H$_2^{13}$CO lines respectively. The arrows represent lower
    limits.}
  \label{fig_modelobs}  
\end{figure}

To illustrate the reliability of the derived jumps,
Fig. \ref{fig_modelobs} shows the ratio between the model and the
observations in the cases of no abundance jump and with a jump, for
\object{NGC1333-IRAS2} as an example.  The constant abundance model
can reproduce the observed fluxes of the lower transitions well, but
it definitively underestimates the flux of the lines originating from
the higher levels.
 
In the next section, we discuss the effects of the other model
parameters on the H$_2$CO abundance determination.
 
\subsection{Dependence on other parameters of the model} 
\label{sec:depend-other-param}

The derived formaldehyde abundances depend on four hidden parameters:
the adopted velocity and density profiles, the H$_2$CO ortho to para
ratio and the evaporation temperature.  In the following we discuss
the influence of all these parameters on the determination of the
H$_2$CO inner abundance.

\subsubsection{Velocity profile}
 

In our analysis, we assumed a velocity profile of a free falling
envelope, given by:
 
\begin{equation} 
v(r) = \left( \frac{2GM}{r} \right)^{1/2} 
\end{equation} 

\noindent
where G is the gravitational constant and M the mass of the central
star. The choice of a free falling velocity profile seems natural, as
these protostars are believed to be in accretion phase
\citep[e.g.][]{Andre00}. Yet, the central mass is a poorly constrained
parameter. Recently, \citet{Maret02} and \citet{Ceccarelli00a} derived
a central mass of 0.5 and 0.8 \solarmass for \object{NGC1333-IRAS4A}
and \object{IRAS16293-2422} respectively. Here we adopt a central mass
of 0.5 \solarmass for all the observed sources, but this parameter
could vary from one source to the other.

A different choice for the  velocity profile could change the derived
abundance. In particular, a higher central mass would imply a higher
velocity gradient in the central parts of the envelope, and as a
consequence, a lower opacity of the H$_2$CO lines. This lower opacity
would decrease the formaldehyde abundance needed to reproduce a given
flux. These differences are expected to affect mainly the high lying
lines, which originate in the inner parts of the envelope.

\begin{figure} 
\centering{
    \includegraphics[width=\hsize]{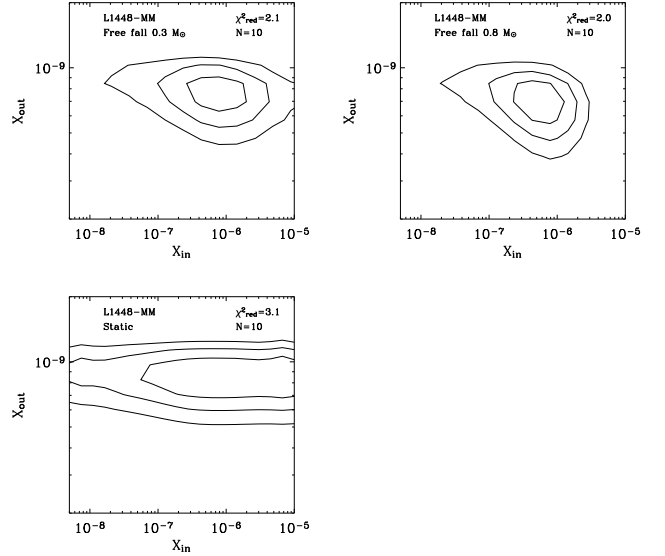}
    \caption{$\chi_{\rm red}^2$ diagrams of \object{L1448-MM} for three
      different velocity profiles. The upper left and upper right
      panels show the $\chi_{\rm red}^2$ contours derived for a free falling
      envelope with a central mass of 0.3 and 0.8 \solarmass
      respectively. The lower panel shows the static envelope case,
      with a 1 km s$^{-1}$ turbulent velocity. Contours indicate the 1, 2
      and 3$\sigma$ confidence levels respectively.}
  }
  \label{fig_velocity_L1448MM}
\end{figure} 

In order to quantify this effect on the derived formaldehyde
abundances, two models of \object{L1448-MM} were run, using a central
mass of 0.3 and 0.8 \solarmass respectively. A third model was also
run to check the case of a static envelope, where a turbulent line
broadening of 1 km s$^{-1}$ has been added. This model was adopted by
\citet{Jorgensen02}, and reproduced well the observed low J CO
emission, which originates in the static envelope.

Figure \ref{fig_velocity_L1448MM} shows the $\chi_{\rm red}^{2}$
diagrams obtained in the three cases. While the derived outer
formaldehyde abundance is not much affected by the adopted velocity
field, the inner abundance changes significantly when adopting a
static envelope rather than a free-fall profile. Yet, the inner
abundance is well constrained in the first two cases (free-fall with
different central masses), and very weakly depends on the adopted
central mass: 8 $\times$ 10$^{-7}$ and 5 $\times$ 10$^{-7}$ for 0.3
and 0.8 \solarmass respectively. On the contrary, only a lower limit
is obtained if a static envelope is adopted: $> 3 \times 10^{-8}$.
This is due to different line opacities in the three cases.  In a
static envelope, the high lying lines become more easily optically
thick, because of the reduced linewidth with respect to the free-fall
cases. For this reason, these lines do not depend on \Xin, when \Xin
is $\sim 10^{-7}$, because they become optically thick.  This explains
why only a lower limit on \Xin can be computed in that case.

\subsubsection{Density profile}


The H$_2$CO abundances depend on the density profile derived from the
continuum observations. In particular, the observations used to derive
the physical structure of the envelopes have been obtained with a
typical beamwidth of $10''$, i.e. 2200 AU at the distance of
NGC1333. The observed continuum is therefore not very sensitive to the
physical conditions in the innermost regions of the envelope, at
scales smaller than a few hundred AUs. Consequently, the derived
density power-law index reflects rather the density in the outer part
of the envelope, and the inner density, extrapolated from these power
law, may be a rough estimate. Finally, the determination of the
density profiles of some of the sources of the sample was difficult
because of the existence of multiple components
\citep{Jorgensen02}. \citet{Jorgensen02} reported an average
uncertainty of $\pm 0.2$ on the power-law index. If the density at the
outer radius of the envelope is assumed to be correctly determined by
the continuum observations, the uncertainty on the power-law index
corresponds to an uncertainty of a factor five on the density at the
inner radius of the envelope. In order to check the effect of this
uncertainty on the derived abundances, we ran models of
\object{NGC1333-IRAS4B} with an inner density artificially multiplied
by a factor 5 (note that the outer density is not changed).  Whereas
the H$_2$CO outer abundance remains unchanged, the inner abundance
decreases by about the same factor 5. Uncertainties in the inner
density could therefore lead to uncertainties on the derived inner
abundances of the same order of magnitude.

\subsubsection{The H$_2$CO ortho to para ratio}
\label{sec:h_2co-ortho-para}


The derived formaldehyde abundances depend also on the value of the
H$_2$CO ortho to para ratio.  Given the relatively low number of
observed lines, this parameter cannot be constrained by the present
observations.  Actually, it is very badly constrained even in the case
of \object{IRAS16293-2422}, where many more formaldehyde lines have
been observed. \cite{Ceccarelli00b} and \citet{Schoier02} report a
value for the ortho and para ratio around 3, but with a large
uncertainty. We thus adopted the canonical value of 3 in our analysis
\citep{Kahane84}.

\begin{figure} 
  \centering{
    \includegraphics[width=\hsize]{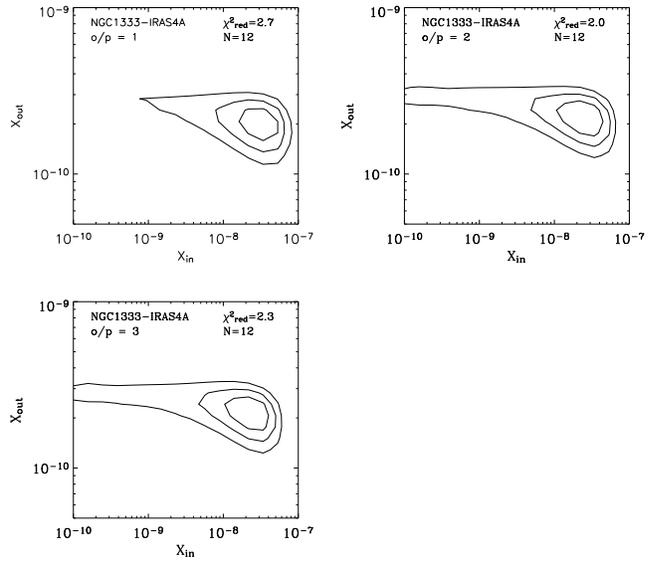}
    \caption{$\chi_{\rm red}^2$ diagrams of \object{NGC1333-IRAS4} for
      three different ortho to para ratios. The upper left, upper
      right and lower panels shows the $\chi_{\rm red}^2$ contours
      derived for an ortho to para ratio of 1, 2 and 3
      respectively. Contours indicates the 1, 2 and 3$\sigma$
      confidence levels respectively.}
    \label{fig_op_IRAS4A}
  }
\end{figure}

As an example, Fig. \ref{fig_op_IRAS4A} shows the influence of this
parameter on the derived H$_2$CO abundance of \object{NGC1333-IRAS4A}.
We ran models with the ortho to para ratio 1, 2 and 3 respectively.
While the derived inner and outer abundances are almost identical for
the three ratios, the abundance jump is slightly better constrained
for a ratio of 1 ($3\sigma$) than an higher ratio ($2\sigma$).  


\subsubsection{Evaporation temperature}

\begin{figure}
  \centering{
    \includegraphics[width=\hsize]{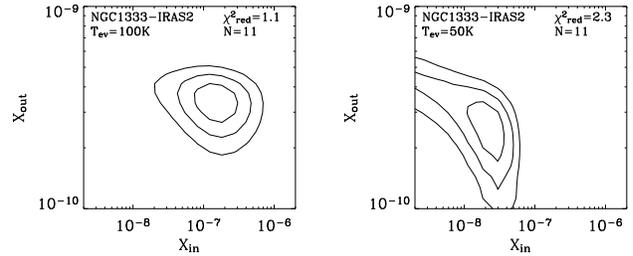}
    \caption{$\chi_{\rm red}^2$ diagrams of \object{NGC1333-IRAS2} for an
      evaporation temperature of 100 K (left panel) and 50 K (right
      panel). Contours indicates the 1, 2 and $3 \sigma$ confidence
      levels respectively.}
    \label{fig_chi2_difftemp}
  } 
\end{figure}

Finally, the evaporation temperature $T_{\rm ev}$, at which the
formaldehyde desorption occurs, is also a hidden parameter of our
model.  As described in \S 5.1, in the present study we adopted
$T_{\rm ev}$ = 100 K, which corresponds to the evaporation of water
ices \citep{Aikawa97}. However, part of the desorption could also
occur at the evaporation temperature of pure formaldehyde ices
\citep[50 K;][]{ Aikawa97,Rodgers03}. For example, a detailed analysis
of the formaldehyde spatial emission in \object{IRAS16293-2422} has
shown that the formaldehyde abundance has a first jump, of about a
factor 10, where $T_\mathrm{dust}$ $\geq$ 50 K, and a second jump of
about a factor 25 where $T_\mathrm{dust}$ $\geq$ 100
\citep{Ceccarelli01}. However, given the relatively small number of
observed lines and the absence of spatial information on the
formaldehyde emission in the source sample of the present study, we
limited the H$_2$CO abundance profile to a single step function.  In
order to check if the data are also consistent with a jump at 50 K, a
model with a jump in the abundance at 50 K for \object{NGC1333-IRAS2}
was run (see Fig. \ref{fig_chi2_difftemp}). While \Xout is very little
sensitive to this parameter, \Xin is about ten times smaller when
assuming a jump at 50 K (2 $\times$ 10$^{-7}$ and 2 $\times$ 10$^{-8}$
for 100 and 50 K respectively). We note, however, that the best
agreement with the data is obtained for an evaporation temperature of
100 K ($\chi^2_{\mathrm{red}}$ = 1.1 against 2.3 respectively).

\subsubsection{Summary}

In the $\chi_{\rm red} ^2$ analysis of \S 5.2, we adopted the most
reasonable values of the various hidden parameters in the model.  In
this section we have explored the effect of variations in them on the
derived abundance jumps.  We emphasize that the actual value of the
jump as well as the precise location are quite uncertain.  Based on
the previous analysis, the adopted velocity field seems to affect most
strongly the amplitude and/or the presence of the derived jump, in
particular when a static, turbulent field is considered.  The other
three parameters, the inner density, the ortho to para ratio, and the
evaporation temperature seem to play a minor role.  In this sense, the
model assumption of an infalling envelope is probably the most
critical in the present analysis.  As already mentioned, we favor the
hypothesis of collapsing envelopes, both because evidences are
accumulating in this direction \citep[e.g.][]{DiFrancesco01}, and
because it is the most natural one.

To summarize, the sources (i.e. \object{NGC1333-IRAS4B},
\object{NGC1333-IRAS2}, \object{L1448-MM} and \object{L1527}) where
the $\chi_{\rm red}^2$ analysis yields $3\sigma$ evidence for jumps,
the presence of an abundance jump is rather firm in our opinion.
Although more marginal, the data are consistent with the presence of a
jump in the other surveyed sources as well.  Appendix
\ref{sec:models-pred-form} lists the predicted fluxes of the brightest
ortho formaldehyde lines. Predictions of para H$_2$CO line fluxes can
be found on the web site
MEPEW\footnote{\url{http://www-laog.obs.ujf-grenoble.fr/~ceccarel/mepew/mepew.html}}
\citep{Ceccarelli03}. In particular, the submillimeter lines are
sensitive to the presence and amplitude of the jump in the H$_2$CO
abundance, and future observations with existing (JCMT, CSO) and
future (e.g. SMA, ALMA) submillimiter telescopes will better constrain
this value.

\section{Discussion} 
 
The first remarkable and by far the most important result of this
study is the evidence for a region of formaldehyde evaporation in
seven out of eight observed sources. In these regions, the
formaldehyde abundance jumps to values two or more orders of magnitude
larger than the abundance in the cold outer envelope. The transition
is consistent with the location where the dust temperature reaches 100
K. The radius of these warm regions is between 13 and 133 AU, and the
densities\footnote{\object{L1527} is an exception, with a density of
$3\times 10^6$ cm$^{-3}$, but, as commented by \citet{Jorgensen02}
this may be due to the contribution of the disk, that may dominate the
continuum emission in the inner parts of this source
\citep{Loinard02b}.}  are between 1 and 20 $\times$ 10$^8$ cm$^{-3}$.
A straightforward interpretation of this result is that the grain
mantles sublimate at 100 K, releasing into the gas phase their
components, and, among them, formaldehyde.  In addition, recent
observations have shown the presence of complex molecules, typical of
massive hot cores \citep{Cazaux03} towards \object{IRAS16293-2422},
the first studied hot core of low mass protostars
\citep{Ceccarelli00a,Ceccarelli00c,Ceccarelli00b,Schoier02}. The
similarity with the well studied hot cores of the massive protostars
is certainly striking: \emph{warm, dense, and compact regions
chemically dominated by the mantle evaporation}.  Even though the
chemistry can be, and very probably is different in high and low mass
protostars, the hot cores represent basically the same process in both
type of sources. Our study does not address the possibility that some
of the H$_2$CO ice mantles may be liberated by shocks interacting with
the inner envelope, since the line wings have been excluded from our
analysis. The role of shocks could be studied by future high angular
resolution maps of the line center and line wing emission.

\subsection{\Xin versus $L_\mathrm{smm}/L_\mathrm{bol}$}
\label{sec:xin-vers-lsmm_lbol}

\begin{figure}
   \centering{
    \includegraphics[width=\hsize]{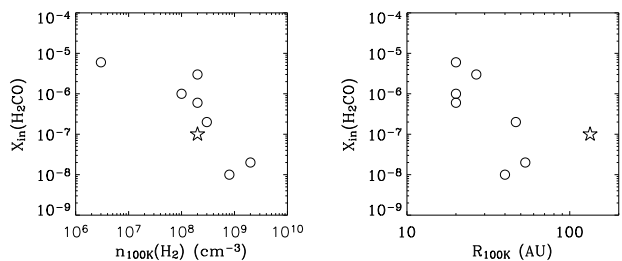}
    \caption{Derived H$_2$CO \Xin abundances as a function of the
      density (left panel) and the radius (right panel) where
      T$_\mathrm{dust}$ = 100 K. The star represent
      \object{IRAS16293-2422}.}
    \label{fig_correl_xin_1}
} 
\end{figure} 

\begin{figure}
   \centering{
    \includegraphics[width=8cm]{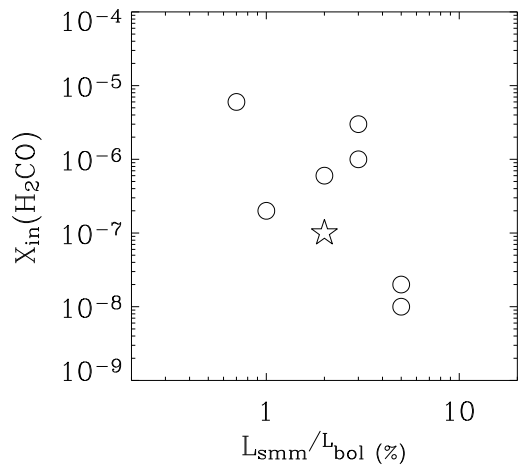}
    \caption{Derived H$_2$CO \Xin abundances as a function of
      $L_\mathrm{smm}/L_\mathrm{bol}$. The star represent
      \object{IRAS16293-2422}.}
    \label{fig_correl_xin_2}
} 
\end{figure} 

The H$_{2}$CO abundance in the inner region shows a loose trend with a
variety of source characteristics.  These include the density in the
inner region and the size of the region (Fig. \ref{fig_correl_xin_1}),
as well as the ratio of the submillimeter luminosity to the bolometric
luminosity, $L_{smm}/L_{bol}$ (Fig. \ref{fig_correl_xin_2}).  In
assessing these correlations, we should, of course, keep the large
uncertainties -- an order of magnitude -- as well as systematic
effects in mind.  In particular, underestimating the density will
immediately result in overestimating the abundance required to explain
the observations (cf. section \ref{sec:depend-other-param}).
Likewise, if we underestimate the size of the emitting region, we will
need a higher H$_2$CO abundance.  Now, which of these, if any,
correlations is the prime correlation and which one is derived is not
directly clear.  In particular, within the inside-out-collapse, the
density and inner radius are anti-correlated as are the luminosity and
the radius of the evaporated region.

The apparent anti-correlation between the H$_{2}$CO abundance and
$L_{smm}/L_{bol}$ (Fig. \ref{fig_correl_xin_2}) is of particular
interest.  The latter is generally taken as an indicator of the
evolutionary stage of the protostar where a larger value for
$L_{smm}/L_{bol}$ implies ``youth'' (e.g. large amounts of cold
material surrounding the YSO).  The anti-correlation may seem a
surprising result, as the most accepted scenario predicts that
formaldehyde is formed on the grain surfaces, likely by successive
hydrogenation of CO \citep{Tielens82,Charnley97} during the
pre-stellar phase. Once in the gas phase because of the evaporation of
the grain mantles, formaldehyde is destroyed (i.e. converted into more
complex molecules) on a timescale of $\sim 10^4$ yr
\citep{Charnley92}.  In this picture, the youngest sources should have
the largest \Xin, which is evidently not the case.  This picture,
however, might be somewhat over-simplified. Indeed, the process of ice
evaporation is continuous, involving larger and larger regions as the
time passes and the luminosity of the protostar increases -- as
pointed out by the models by \citet{Ceccarelli96} and
\citet{Rodgers03} -- so that the result is a shell of continuously
refurbished H$_2$CO, moving outwards.  The main point is that the
region of formaldehyde overabundance never disappears, but just moves,
and the jump in the H$_2$CO abundance is only governed by the quantity
of formaldehyde in the grain mantles.

If the $L_\mathrm{smm}/L_\mathrm{bol}$ ratio is not an age indicator
but rather a parameter affected more by the different initial
conditions of each protostar, and specifically it is larger in sources
where the pre-stellar density is larger
\citep[e.g.][]{Jayawardhana01}, the trend of Fig. \ref{fig_correl_xin_2}
would imply that the H$_2$CO ice abundance depends on the pre-stellar
conditions.  Less dense pre-stellar conditions (i.e. lower
$L_\mathrm{smm}/L_\mathrm{bol}$ ratios) would give rise to more
H$_2$CO enriched ices, because there is more atomic H and thus more
grain surface hydrogenation to form H$_2$CO.  This is indeed
consistent with the fact that the efficiency of CO hydrogenation into
H$_2$CO on the grain mantles is $\sim$ 250 times larger in H$_2$O-rich
ices when compared to CO-rich ices \citep{Ceccarelli01}.  And since
less dense regions have likely more H$_2$O-rich than CO-rich ices,
because CO-rich ices likely form in relatively dense condensations
(Bacmann et al. 2002), the larger H$_2$CO abundance in sources with a
lower $L_\mathrm{smm}/L_\mathrm{bol}$ would therefore be due to a real
larger efficiency of the H$_2$O-rich ices in forming H$_2$CO.  As a
consequence, our finding would suggest that the
$L_\mathrm{smm}/L_\mathrm{bol}$ ratio does not probe the evolutionary
status of protostars, but rather mainly reflects their different
initial conditions.

Alternative explanations are also possible.  For example, recent
laboratory works suggest that the formation of formaldehyde by CO
hydrogenation on the grains depends on the dust temperature
\citep{Watanabe03}, and this may also be consistent with ``older''
protostars (i.e. lower $L_\mathrm{smm}/L_\mathrm{bol}$ ratios), being
also hotter, having larger H$_2$CO abundances. This, of course, would
imply that the bulk of the H$_2$CO is formed in a stage later than the
CO condensation, namely during the pre-stellar core phase
\citep{Bacmann02}.  Whether this is likely is difficult to say, for CO
may indeed be trapped on the grain mantles and partly converted into
H$_2$CO only when the grain temperature increases, as suggested by the
laboratory experiments.  Another possibility is that formaldehyde
formation on grains needs UV radiation
\citep[e.g][]{dHendecourt86,Schutte96}. Analogously to above,
formaldehyde would be formed only in a later stage, and the older the
protostar, the larger the UV field and the larger the H$_2$CO
abundance.

All these interpretations need to be taken with caution, of course,
since the inferred variations in the inner H$_{2}$CO abundance might
reflect uncertainties in the density and/or size of the region
emitting the H$_{2}$CO lines.  A similar study on a larger sample and
focussing on higher energy lines is required to draw more definitive
conclusions.

\subsection{Low versus high mass protostars}

In section \ref{envelope_model} we have examined the evidence for the
presence of jumps in the H$_2$CO abundance in the warm gas surrounding
low mass YSOs.  The presence of such abundance jumps in hot cores
around high mass stars is not well established.  On the one hand, the
prototype of hot cores in regions of massive star formation -- the hot
core in the Orion BN/KL region -- has a H$_2$CO abundance of $10^{-7}$
\citep{Sutton95}.  On the other hand, in a study of hot cores in a
sample of massive protostars, \citet{vanderTak00} did not find
evidence for the presence of H$_2$CO abundance jumps, but did find
evidence for jumps in the CH$_3$OH abundance.  It is unclear at
present whether the Orion hot core or the van der Tak sample is more
representative for the composition of hot cores in regions of massive
star formation.  Presuming that the differences in H$_2$CO abundance
jumps are real, we note that the composition of the ices -- which
drive the chemistry in hot cores -- may well differ between regions of
low mass and high mass star formation.

Supporting this idea, the deuterium fractionation is dramatically
different in the high and low mass protostars.  Doubly deuterated
formaldehyde and methanol have been observed to be extremely abundant
in low mass protostars when compared to massive protostars.  The
D$_2$CO/H$_2$CO ratio is more than 25 times larger in low than in high
mass protostars \citep{Ceccarelli98,Loinard02a}.  Deuterated methanol
may be as abundant than its main isotopomer in the low mass protostar
\object{IRAS16293-2422} \citep{Parise02}, whereas it is only 4\% of
CH$_3$OH in Orion \citep{Jacq93}.  Since this extreme deuteration is
likely a grain mantle product \citep[e.g.][]{Ceccarelli01,Parise02},
the dramatic differences in the molecular deuteration between low and
high mass protostars are already a very strong indication that mantles
in both type of sources are \emph{substantially} different.  This
indeed does not comes as a surprise, as the pre-collapse conditions
very likely differ: warmer in high than in low mass stars, at the very
least.


\subsection{\Xout versus CO abundance}

\begin{figure}
  \centering{
    \includegraphics[width=8cm]{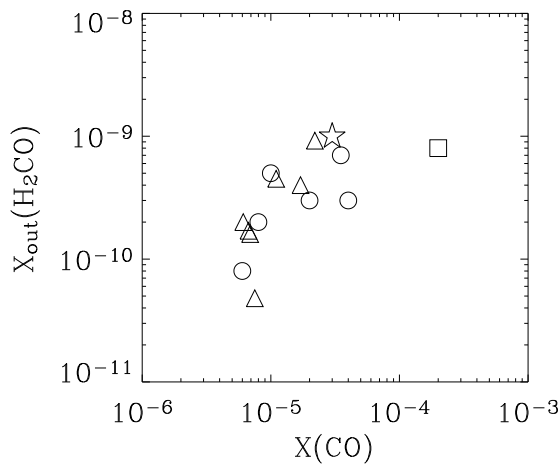} 
    \caption{Derived outer formaldehyde abundance \Xout as a function
      the CO abundance of the outer enveloppe.  Circles represent
      Class 0 sources, The star represents \object{IRAS16293-2422},
      the square represents \object{VLA1623}, and the triangles
      represent the pre-stellar cores of \citet{Bacmann02,Bacmann03}
      sample.}
    \label{fig_correl_xout} 
  }
\end{figure}

Finally, Fig. \ref{fig_correl_xout} compares the H$_2$CO abundance
\Xout with the CO abundance derived by \citet{Jorgensen02}, in the
outer envelope. On the same plot we also reported the values found in
the prestellar cores studied by \citet{Bacmann02,Bacmann03}. The first
thing to notice is the similarity of the values found in Class 0
sources and pre-stellar cores, in both molecules, despite the
different methods used to derive the abundances.  The similarity of
the values adds support to the thesis that the pre-stellar cores of
the Bacmann et al. sample are precursors of Class 0 sources, and that
the outer regions of the envelopes of Class 0 sources are formed by
material still unchanged by the collapse, i.e. that reflects the
pre-collapse conditions.  Second aspect to note of
Fig. \ref{fig_correl_xout} is the clear correlation between the
H$_2$CO and CO abundance.  In this case the interpretation is
straightforward: in the outer, cold envelope molecules are depleted,
and the degree of depletion is the same for the CO and the H$_2$CO
molecules.  As discussed in \citet{Bacmann02}, the limited CO
depletion observed in pre-stellar cores strongly suggests that a
relatively efficient mechanism (cosmic rays?) re-injects CO into the
gas phase.  Since the binding energies of the CO and H$_2$CO are
relatively different \citep[e.g.][]{Aikawa97}, the linear correlation
of Fig. \ref{fig_correl_xout} strengthens the claim that H$_2$CO
molecules are trapped into CO-rich ices \citep{Ceccarelli01}.



\section{Conclusions} 
 
We have presented a spectral survey of the formaldehyde emission of a
sample of eight Class 0 protostars, carried out with the IRAM 30m and
JCMT telescopes. A total of eight transitions were selected for each
source, covering a large range of upper level energies in order to
probe different physical conditions. When possible, the isotopic lines
were observed in order to derive the line opacities. Most of the lines
are relatively narrow with a small contribution from wings extending
to larger velocities.  Using the standard rotational diagram method,
we derived rotational temperatures between 11 and 40 K, and H$_{2}$CO
column densities between 1 $\times$ 10$^{13}$ and 7 $\times$ 10$^{13}$
cm$^{-2}$.  For the sources with detected H$_{2}^{13}$CO lines,
opacity corrections increase the derived column densities to the range
0.8 and 2 $\times 10^{14}$ cm$^{-2}$. In order to test the effect of
non-LTE excitation, the observed emission has been modeled using a LVG
code.  In this way the derived temperatures are significantly higher
than the rotational temperatures, suggesting that the observed
transitions are sub-thermally populated. The inferred densities,
between 1 and 6 $\times$ 10$^{5}$ cm$^{-3}$, indeed support this
conclusion.

To take into account the density and temperature gradients in the
protostellar envelopes, the emission has been modeled using densities
and dust temperature profiles derived from previous studies of the
continuum emission of these objects.  The gas temperature in the
envelopes was computed using a code of the thermal balance for
protostellar envelopes.  The formaldehyde abundance across the
envelope has been approximated by a step function : an outer abundance
\Xout where $T_{\rm dust} \leq$ 100 K, and a inner abundance \Xin at
$T_{\rm dust} \geq$ 100 K.  We show that the outer abundance, \Xout,
is well constrained in all the sources, and varies between 8 $\times$
10$^{-11}$ and 8 $\times$ 10$^{-10}$.  The inner abundance \Xin is
well constrained in four sources with a 3 $\sigma$ level confidence,
while in three sources it is only a $\leq 2 ~ \sigma$ evidence, and no
evidence of a jump is found in \object{VLA1623}.  The derived values
of \Xin range between 1 $\times$ 10$^{-8}$ and 6 $\times$ 10$^{-6}$,
showing jumps of the formaldehyde abundance between 2 and 4 orders of
magnitude.  \emph{The most important conclusion of this study is hence
that large amounts of formaldehyde are injected into the gas phase
when the grain mantles evaporate at 100 K.}

We have discussed the uncertainties on the actual values of the hidden
parameters of the model, namely the velocity and density profile, the
H$_2$CO ortho to para ratio, and the evaporation temperature. The
uncertainty in these parameters makes the abundance jump value and
jump locations uncertain for some sources. Future observations of
higher frequency lines and modeling of the line profiles may
distinguish between the different interpretations.

We found that sources with lower $L_\mathrm{smm}/L_\mathrm{bol}$
ratios possess the largest inner H$_2$CO abundances.  We discussed why
we think that this reflects very likely a different H$_2$CO enrichment
of the grain mantles.


We found that the inner H$_2$CO abundances are systematically larger
than the H$_2$CO abundances of the hot cores of the sample of massive
protostars studied by \citet{vanderTak00}. This supports to the idea
that low and high mass protostars have a different grain mantle
composition.

Finally, the derived outer H$_2$CO abundance shows a clear correlation
with the CO abundance, implying that both molecules are depleted by a
similar factor in the outer envelope, namely that H$_2$CO molecules
are likely trapped in CO-rich ices in the outer envelope.

\acknowledgements{Most of the computations presented in this
  paper were performed at the Service Commun de Calcul Intensif de
  l'Observatoire de Grenoble (SCCI). Astrochemistry in Leiden is
  supported by a NOVA Network 2 PhD grant and by a NWO Spinoza grant.}
 
\bibliography{h2co} 
\bibliographystyle{aa} 

\Online 
\appendix

\section{Density and temperature profile of \object{L1448-N}}
\label{sec:dens-temp-prof}

\begin{figure*}  
    \centering{  
      \includegraphics[width=17cm]{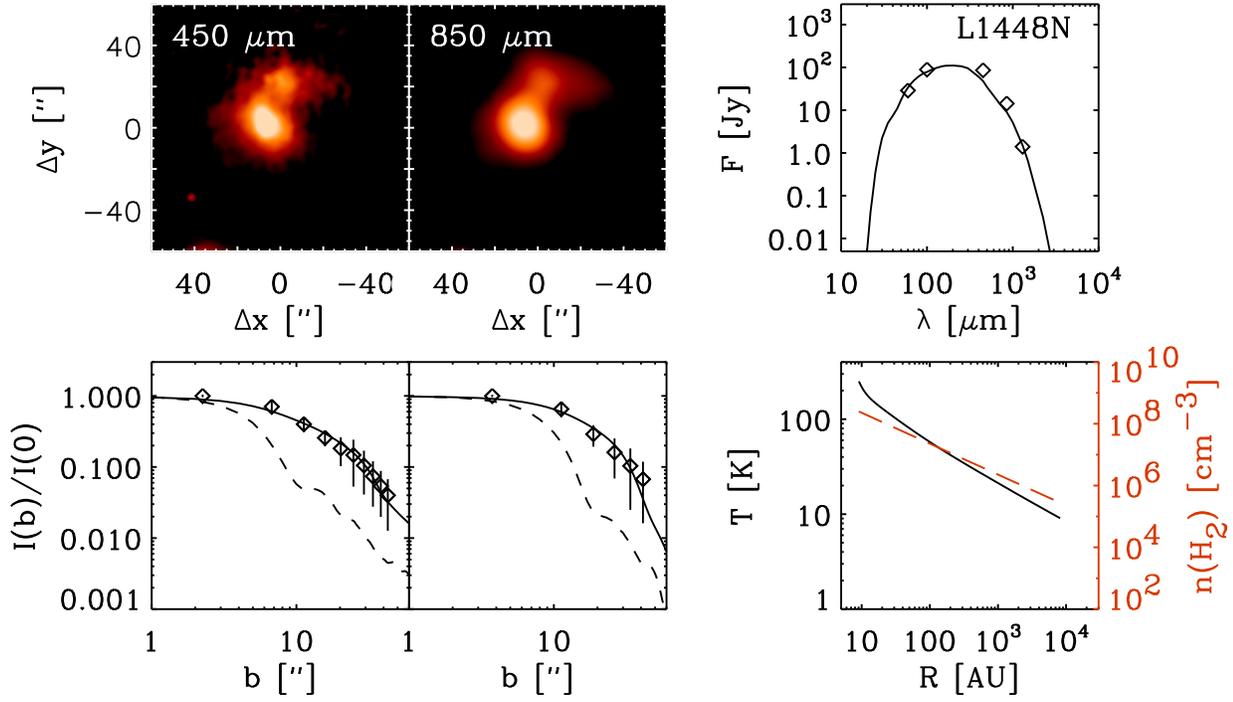} 
    }  
    \caption{Observations and result of the modelling of L1448N. The
    upper left panel shows the SCUBA observations at 450 and 850
    $\mu$m. The upper right panel shows the SED (diamonds) and the
    fitted model (solid line). The lower left panel shows the
    brightness profile and the fitted model. The dashed line represents
    the beam profile. The lower right panel shows the derived
    temperature dust (solid line) and density profile (dashed line).}
    \label{fig_model_l1448n}  
\end{figure*}

The density and temperature profile of \object{L1448-N} have been
determined following the method used by \cite{Jorgensen02}. The
spectral energy distribution (SED) and JCMT-SCUBA maps at 450 and 800
$\mu$m have been compared to the prediction of the radiative code
DUSTY\footnote{DUSTY is publicly available at
\url{http://www.pa.uky.edu/~moshe/dusty/}} \citep{Ivezic97}. The JCMT
observations were taken from the JCMT
archive. Fig. \ref{fig_model_l1448n} shows the result of the fits of
the brightness profile and SED of this source. The envelope parameters
from this modeling are summarized in Table
\ref{table_result_model_l1448n}.
 
\input{table_result_model_l1448n} 
 
\object{L1448-N} shows a relatively flat density profile. This profile may 
reflect the asymmetry of the source and the extension of the 
emission, which can flatten the derived profile \citep[see][ for a 
discussion of the effects of asymmetries in the derived density 
profile]{Jorgensen02}. 
 
\section{Models predictions for formaldehyde lines fluxes}
\label{sec:models-pred-form}

In this Appendix we present the model predictions for the fluxes of
selected o-H$_{2}$CO transitions. In Table \ref{tab:flux_const} we
report the line fluxes computed assuming a constant abundance \Xout
across the envelope, while Table \ref{tab:flux_jump} reports the
fluxes predicted assuming an abundance \Xout in the outer part of the
envelope where $T_\mathrm{dust} < 100$ K, increasing to \Xin in the
inner part of the envelopes, at the radius where $T_\mathrm{dust} \ge
100 K$. The adopted values of \Xin and \Xout are the best fit values
reported in \ref{table_result_infall}. In these two tables, the line
fluxes are expressed in $\mathrm{erg} \ \mathrm{s}^{-1} \
\mathrm{cm}^{-2}$. Although a conversion in $\mathrm{K} \ \mathrm{km}
\ \mathrm{s}^{-1}$ would have been more practical to compare these
fluxes with observations, this conversion depends on the telescope
used and the extent of the line emission. However if the emitting
region is smaller than the telescope beam, the fluxes can be
approximatively converted into $\mathrm{K} \ \mathrm{km} \
\mathrm{s}^{-1}$ using the conversion factors reported by
\cite{Ceccarelli03}.

While low lying lines fluxes are comparable in two tables, higher
energy lines fluxes substantially differ. Higher frequency
observations can therefore help to distinguish between a constant
abundance in the envelope and a jump in the abundance, especially on
the sources of our sample for which the jump in the abundance is
uncertain.

\input{table_flux_const.tex}

\input{table_flux_jump.tex}

\end{document}

%% file: table_sources.tex





\begin{table*}
  \caption{The observed sample. \object{IRAS16293-2422}, which has
  been studied elsewhere (see text) is included for comparison.}
  \centering{
    \begin{tabular}{l l l l l l l l l}
      \hline
      \hline
      Source & $\alpha (2000)$ & $\delta (2000)$ & Cloud &
      Dist.$^{\mathrm{a}}$ & $L_\mathrm{bol}$$^{\mathrm{b}}$ &
      $M_\mathrm{env}$$^{\mathrm{b}}$ & $L_\mathrm{smm} /
      L_\mathrm{bol}$$^{\mathrm{c}}$ & $T_\mathrm{bol}$$^{\mathrm{b}}$ \\
      &&&& (pc) & (\solarlum) & (\solarmass) & (\%) & (K)\\  
      \hline
      \object{NGC1333-IRAS4A} & 03:29:10.3 & +31:13:31 & Perseus & 220
      & 6 & 2.3 & 5 & 34\\
      \object{NGC1333-IRAS4B} & 03:29:12.0 & +31:13:09 & Perseus & 220
      & 6 & 2.0 & 3 & 36\\
      \object{NGC1333-IRAS2} & 03:28:55.4 & +31:14:35 & Perseus & 220
      & 16 & 1.7 & $\lesssim$ 1 & 50\\
      \object{L1448-MM} & 03:25:38.8 & +30:44:05 & Perseus & 220 & 5 &
      0.9 & 2 & 60\\
      \object{L1448-N} & 03:25:36.3 & +30:45:15 & Perseus & 220 & 6 &
      3.5 & 3 & 55\\
      \object{L1157-MM} & 20:39:06.2 & +68:02:22 & Isolated & 325 & 11
      & 1.6 & 5 & 60\\
      \object{L1527} & 04:39:53.9 & +26:03:10 & Taurus & 140 & 2 & 0.9
      & 0.7 & 60\\
      \object{VLA1623} & 16:26:26.4 & -24:24:30 & $\rho$-Ophiuchus &
      160 &  1 & 0.2 & 10 & $<$ 35\\ \hline
      \object{IRAS16293-2422}$^{\mathrm{d}}$ & 16:32:22.7 & -24:38:32 &
      $\rho$-Ophiuchus & 160 & 27 & 5.4 &  2 & 43\\
      \hline
    \end{tabular}
    \begin{itemize}
      \item[$^{\mathrm{a}}$]{From \citet{Andre00}, except for Perseus
	sources \citep{Cernis90}.}
      \item[$^{\mathrm{b}}$]{From \citet{Jorgensen02}}
      \item[$^{\mathrm{c}}$]{From \citet{Andre00}}
      \item[$^{\mathrm{d}}$]{Included for comparison.}
    \end{itemize}	
  }
  \label{sources}
\end{table*}

%% file: table_flux_h2co.tex
\begin{table*}
  \begin{center}
    \caption{Integrated fluxes of H$_{2}$CO lines in $T_\mathrm{mb} \,
      \mathrm{d}V$ units . Upper limits are given as 2 $\sigma$. The
      ``-'' symbol indicates lack of the relevant observation.}
    \begin{tabular}{l l l l l l l l l l}
      \hline
      \hline
      &\multicolumn{3}{c}{o-H$_2$CO}&&\multicolumn{5}{c}{p-H$_2$CO}\\
      \cline{2-4}\cline{6-10}
      Transition & 2$_{1,2}$-1$_{1,1}$ &
      4$_{1,4}$-3$_{1,3}$ & 5$_{1,5}$-4$_{1,4}$ & & 3$_{0,3}$-2$_{0,2}$ & 
      3$_{2,2}$-2$_{2,1}$ & 5$_{0,5}$-4$_{0,4}$ &
      5$_{2,4}$-4$_{2,3}$ & 5$_{4,2}$-4$_{4,1}$$^{\mathrm{a}}$\\
      \hline
      $E_{u}$ (K)& 21.9 & 45.6 & 62.4 & & 21.0 & 68.1 & 52.2 &
      96.7 & 234\\
      $A_{u,l}$ (s$^{-1}$) & 5.4 $\times$ 10$^{-5}$ &
      6.0 $\times$ 10$^{-4}$ &
      1.2 $\times$ 10$^{-3}$ & & 2.9 $\times$ 10$^{-4}$ &
      1.6 $\times$ 10$^{-4}$ & 1.4 $\times$ 10$^{-3}$ &
      1.2 $\times$ 10$^{-3}$ & 5.0 $\times$ 10$^{-4}$\\
      $\nu$ (GHz) & 140.839 & 281.527 &
      351.769 & & 218.222 & 218.476 & 362.736 & 363.946 & 364.103\\
      $g_u$ & 5 & 9 & 11 & & 7 & 7 & 11 & 11 & 11\\
      Telescope & IRAM & IRAM & JCMT & & IRAM & IRAM & JCMT & JCMT &
      JCMT\\
      HPBW ('') & 17 & 9 & 14 & & 11 & 11 & 14 & 14 & 14\\
      $\eta_\mathrm{mb}$ or $B_\mathrm{eff}$/$F_\mathrm{eff}$ & 0.74 &
      0.47 & 0.63 && 0.62 & 0.62 & 0.63 & 0.63 & 0.63\\
      \hline
      \object{NGC1333-IRAS4A} & 9.1 $\pm$ 1.4 & 10.6 $\pm$ 2.6 & 5.5
      $\pm$ 1.2 & & 9.3 $\pm$ 1.9 & 2.2 $\pm$ 0.4 & 2.9 $\pm$ 1.0 
      & 1.2 $\pm$ 0.6 &  1.7 $\pm$ 0.9\\
      \object{NGC1333-IRAS4B} & 6.8 $\pm$ 1.0 & 12.1 $\pm$ 3.0 & 7.5
      $\pm$ 1.7 & & 9.6 $\pm$ 1.9 & 4.7 $\pm$ 1.0 & 5.9 $\pm$ 2.1 
      & 3.6 $\pm$ 1.0 & 0.9 $\pm$ 0.6\\
      \object{NGC1333-IRAS2} & 4.3 $\pm$ 0.6 & 5.8 $\pm$ 1.5 & 1.6
      $\pm$ 0.4 & & 4.9 $\pm$ 1.0 & 1.0 $\pm$ 0.2 & 1.8 $\pm$ 0.6 
      & 0.6 $\pm$ 0.4 & $<$ 0.4\\
      \object{L1448-MM} & 3.3 $\pm$ 0.7 & 4.7 $\pm$ 1.1 & 1.0 $\pm$
      0.2 & & 3.4 $\pm$ 0.6 & 0.4 $\pm$ 0.1 & 1.3 $\pm$ 0.4 & 0.2
      $\pm$ 0.1 & $<$ 0.1\\
      \object{L1448-N} & 7.9 $\pm$ 0.9 & 6.9 $\pm$ 1.6 & 3.8 $\pm$ 0.9
      & & 5.7 $\pm$ 0.9 & - & - & 0.6 $\pm$ 0.2 & - \\
      \object{L1157-MM} & 1.2 $\pm$ 0.2 & 2.1 $\pm$ 0.5 & 1.2 $\pm$
      0.3 & & 1.1 $\pm$ 0.3 & $<$ 0.2 & 0.5 $\pm$ 0.2 & $<$ 0.1 & $<$
      0.1\\
      \object{L1527} & 2.8 $\pm$ 0.7 & 4.5 $\pm$ 1.1 & 1.0 $\pm$ 0.3 &
      & 3.0 $\pm$ 1.3 & 0.2 $\pm$ 0.1 & 0.4 $\pm$ 0.2 & - & -\\  
      \object{VLA1623} & 3.8 $\pm$ 1.2 & 2.7 $\pm$ 1.2 & 0.9 $\pm$ 0.2
      & & 5.0 $\pm$ 1.5 & - & 1.2 $\pm$ 0.4 & $<$ 0.2 & $<$ 0.3\\
      \hline
   \end{tabular}
    \begin{itemize}
      \item[$^{\mathrm{a}}$]{Blended with the 5$_{4,1}$-4$_{4,0}$
      H$_{2}$CO line.}
    \end{itemize}
  \label{table_fluxes_h2co}
 \end{center}
\end{table*}


%% file: table_flux_h213co.tex






\begin{table*}
  \begin{center}
    \caption{Integrated fluxes of H$_{2}$$^{13}$CO lines in
      $T_{\mathrm mb} \, \mathrm{d}V$ units. Upper limits are given as 2
      $\sigma$. The ``-'' symbol indicates lack of the relevant
      observation.}
    \begin{tabular}{l l l l l l}
      \hline
      \hline
      &\multicolumn{3}{c}{o-H$_2^{13}$CO} & &
      \multicolumn{1}{c}{p-H$_2^{13}$CO}\\
      \cline{2-4}\cline{6-6}
      Transition & 2$_{1,2}$-1$_{1,1}$ &
      4$_{1,4}$-3$_{1,3}$ & 5$_{1,5}$-4$_{1,4}$ & &
      5$_{2,4}$-4$_{2,3}$ \\
      \hline
      $E_{u}$ (K)& 21.7 & 44.8 & 61.3 & & 98.5\\ $A_{u,l}$ (s${-1}$) &
      1.5 $\times$ 10$^{-4}$ & 1.7 $\times$ 10$^{-3}$ & 3.4 $\times$
      10$^{-3}$ & & 1.1 $\times$ 10$^{-3}$\\
      $\nu$ (GHz) & 137.450 & 274.762 &
      343.325 & & 354.899\\
      HPBW ('') & 17 & 9 & 14 & & 14\\
      Telescope & IRAM & IRAM & JCMT & & JCMT\\
      $\eta_\mathrm{mb}$ or $B_\mathrm{eff}$/$F_\mathrm{eff}$ & 0.74 &
      0.47 & 0.63 & & 0.63\\
      \hline
      \object{NGC1333-IRAS4A} & 0.4 $\pm$ 0.1 & 0.3 $\pm$ 0.1 &
      0.3 $\pm$ 0.2 & & $<$ 0.1\\
      \object{NGC1333-IRAS4B} & - & - & 0.3 $\pm$ 0.1 & & $<$ 0.1\\
      \object{NGC1333-IRAS2} & 0.2 $\pm$ 0.1 & - & 0.4 $\pm$ 0.2 & &
      $<$ 0.1\\
      \object{L1448-MM} & - & - & $<$ 0.1 & & $<$ 0.2\\
      \object{L1448-N} & 0.2 $\pm$ 0.1 & $<$ 0.1 & - & & -\\
      \object{L1157-MM} & - & - & $<$ 0.1 & & $<$ 0.1\\
      \hline
   \end{tabular}
    \label{table_fluxes_h213co}
  \end{center}
\end{table*}


%% file: table_opacities.tex


\begin{table*}
  \begin{center}
    \caption{H$_{2}$CO lines opacities derived from the H$_{2}$$^{13}$CO
      observations.}
    \begin{tabular}{l l l l l l}
      \hline
      \hline
      &\multicolumn{3}{c}{o-H$_2$CO} & & \multicolumn{1}{c}{p-H$_2$CO}\\
      \cline{2-4}\cline{6-6}
      Transition & 2$_{1,2}$-1$_{1,1}$ &
      4$_{1,4}$-3$_{1,3}$ & 5$_{1,5}$-4$_{1,4}$ &
      &5$_{2,4}$-4$_{2,3}$ \\
      \hline
      \object{NGC1333-IRAS4A} & 1.0$^{+0.7}_{-0.4}$ &
      0.5$^{+1.2}_{-0.4}$ & 1.2$^{+9.9}_{-0.7}$ && $<$ 2\\
      \object{NGC1333-IRAS4B} & - & - & 0.9$^{+1.3}_{-0.5}$ && $<$
      0.5\\
      \object{NGC1333-IRAS2} & 1.1$^{+1.9}_{-0.6}$ & - &
      5.5$^{+11.1}_{-2.2}$ && $<$ 4\\
      \object{L1448-MM} & - & - & $<$ 2 && - \\
      \object{L1448-N} & 0.4$^{+1.1}_{-0.3}$ & $<$ 0.1 & - && - \\
      \object{L1157-MM} & - & - & $<$ 1.4 && - \\
      \hline
    \end{tabular}
    \label{table_opacities}
  \end{center}  
\end{table*}


%% file: table_result_LTE_LVG.tex
\begin{table*}
  \begin{center}
    \caption{Results of the rotational diagram and LVG analysis.}
    \label{table_results_LTE_LVG}
    \begin{tabular}{l l l l l l l l l l}
      \hline
      \hline
      &\multicolumn{3}{c}{Rotational Diagram} &&
      \multicolumn{5}{c}{LVG}\\
      \cline{2-4}\cline{6-10}
      Source & $T_\mathrm{rot}$ & $N_\mathrm{thin}$(H$_2$CO) &
      $N$(H$_2$CO)$^\mathrm{a}$ && $T_\mathrm{gas}$ & $n$(H$_2$) &
      $N$(H$_2$CO)$^\mathrm{b}$ & H$_2$CO/H$_2$$^\mathrm{c}$ &
      CO/H$_2$$^\mathrm{d}$ \\
      & (K) & (cm$^{-2}$) & (cm$^{-2}$) && (K) &
      (cm$^{-3}$) & (cm$^{-2}$) &\\
      \hline
      \object{NGC1333-IRAS4A} & 25 & 7 $\times$ 10$^{13}$ &
      2 $\times$ 10$^{14}$ && 50 & 3 $\times$ 10$^{5}$ &
      1 $\times$ 10$^{15}$ & 5 $\times$ 10$^{-10}$ & 8 $\times$
      10$^{-6}$\\
      \object{NGC1333-IRAS4B} & 40 & 7 $\times$ 10$^{13}$ &
      2 $\times$ 10$^{14}$ && 80 & 3 $\times$ 10$^{5}$ &
      2 $\times$ 10$^{14}$ & 7 $\times$ 10$^{-10}$ & 1 $\times$
      10$^{-5}$\\
      \object{NGC1333-IRAS2} & 24 & 3 $\times$ 10$^{13}$ &
      1 $\times$ 10$^{14}$ && 70 & 3 $\times$ 10$^{5}$ &
      5 $\times$ 10$^{13}$ & 1 $\times$ 10$^{-10}$ & 2 $\times$
      10$^{-5}$\\
      \object{L1448-MM} & 19 & 2 $\times$ 10$^{13}$ & - && 30 & 3
      $\times$ 10$^{5}$ & 6 $\times$ 10$^{13}$ & 4 $\times$
      10$^{-10}$ & 4 $\times$ 10$^{-5}$\\ 
      \object{L1448-N} & 22 & 5 $\times$ 10$^{13}$ & 8 $\times$
      10$^{13}$ && 90 & 1 $\times$ 10$^{5}$ & 3 $\times$ 10$^{13}$ & 7
      $\times$ 10$^{-10}$ & - \\
      \object{L1157-MM} & 18 & 1 $\times$ 10$^{13}$ & - && 80 & 3
      $\times$ 10$^{5}$ & 4 $\times$ 10$^{13}$ & 3 $\times$
      10$^{-11}$ & 6 $\times$ 10$^{-6}$\\
      \object{L1527} & 16 & 3 $\times$ 10$^{13}$ & - && 30 & 6
      $\times$ 10$^{5}$ & 4 $\times$ 10$^{13}$ & 1 $\times$
      10$^{-9}$ & 4 $\times$ 10$^{-5}$\\
      \object{VLA1623} & 11 & 3 $\times$ 10$^{13}$ & - && 80 & 1
      $\times$ 10$^{5}$ & 8 $\times$ 10$^{13}$ & 3 $\times$
      10$^{-10}$ & 2 $\times$ 10$^{-4}$\\
      \hline
    \end{tabular}
    \begin{itemize}
    \item[$^\mathrm{a}$]{Corrected for opacity effects, assuming a
    value of $\tau = 1$ for \object{NGC1333-IRAS4A},
    \object{NGC1333-IRAS4B} and \object{NGC1333-IRAS2} and $\tau =
    0.4$ for \object{L1448-N} respectively (see Table \ref{table_opacities}}.
    \item[$^\mathrm{b}$]{Averaged over a 10'' beam.}
    \item[$^\mathrm{c}$]{From H$_2$ column densities quoted by 
      \citet{Jorgensen02}}
    \item[$^\mathrm{d}$]{From \citet{Jorgensen02}}
    \end{itemize}
  \end{center}
\end{table*}

%% file: table_result_infall.tex

\begin{table*}
  \begin{center}
    \caption{Formaldehyde abundances as derived from the envelope
      model in the outer parts of the envelope (\Xout) and the inner
      parts of the envelope (\Xin). The table also includes the radius
      where the dust temperature is 100 K and 50 K, and the density at
      the radius where the dust temperature is 100 K.}
    \label{table_result_infall}
    \begin{tabular}{l l l l l l l}
      \hline
      \hline
      Source & $R_\mathrm{100 K}$ & $R_\mathrm{50 K}$ &$n_\mathrm{100
      K}$ & \Xout & \Xin\\
      & (AU) & (AU) & (cm$^{-3}$) & & \\
      \hline
      \object{NGC1333-IRAS4A} & 53 & 127 & 2 $\times$ 10$^{9}$ & 2
      $\times$ 10$^{-10}$ & 2 $\times$ 10$^{-8}$\\
      \object{NGC1333-IRAS4B} & 27 & 101 & 2 $\times$ 10$^{8}$ & 5
      $\times$ 10$^{-10}$ & 3 $\times$ 10$^{-6}$\\
      \object{NGC1333-IRAS}2 & 47 & 153 & 3 $\times$ 10$^{8}$ & 3
      $\times$ 10$^{-10}$ & 2 $\times$ 10$^{-7}$\\
      \object{L1448-MM} & 20 & 89 & 2 $\times$ 10$^{8}$ & 7 $\times$
      10$^{-10}$ & 6 $\times$ 10$^{-7}$\\
      \object{L1448-N} & 20 & 95 & 1 $\times$ 10$^{8}$ & 3 $\times$
      10$^{-10}$ & 1 $\times$ 10$^{-6}$\\
      \object{L1157-MM} & 40 & 105 & 8 $\times$ 10$^{8}$ & 8 $\times$
      10$^{-11}$ & 1 $\times$ 10$^{-8}$\\
      \object{L1527} & 20 & 140 & 3 $\times$ 10$^{6}$ & 3 $\times$
      10$^{-10}$ & 6 $\times$ 10$^{-6}$\\
      \object{VLA1623} & 13 & 48 & 2 $\times$ 10$^{8}$ & 8 $\times$
      10$^{-10}$ & -\\ 
      \hline
      \object{IRAS16293-2422}$^{\mathrm{a}}$ & 133 & 266 & 1 $\times$
      10$^{8}$ & 1 $\times$ 10$^{-9}$ & 1 $\times$ 10$^{-7}$\\
      \hline
    \end{tabular}
    \begin{itemize}
      \item[$^{\mathrm{a}}$]{From \citet{Ceccarelli00b}.}
    \end{itemize}
  \end{center}
\end{table*}

%% file: table_result_model_l1448n.tex

\begin{table}
  \begin{center}
    \caption{L1448-N best fit parameters from the DUSTY modelling and derived
      physical parameters.}
    \label{table_result_model_l1448n}
    \begin{tabular}{l l}
      \hline
      \hline
      \multicolumn{2}{c}{DUSTY parameters}\\
      \hline
      Y & 1100\\
      $\alpha$ & 1.2\\
      $\tau_{100}$ & 0.8\\
      \hline
      \multicolumn{2}{c}{Derived physical parameters}\\
      \hline
      R$_i$ & 10 AU\\
      R$_\mathrm{10K}$ & 7000 AU\\
      n$_{R_i}$ & 4.9 $\times$ 10$^8$ cm$^{-3}$\\
      n$_\mathrm{10K}$ & 1.1 $\times$ 10$^5$ cm$^{-3}$\\
      M$_\mathrm{10K}$ & 2.5 M$_{\sun}$\\
      \hline
    \end{tabular}
  \end{center}
\end{table}

%% file: table_flux_const.tex
\begin{table*}
  \caption{Predicted line fluxes of selected o-H$_2$CO transitions for
    a constant abundance \Xout across the enveloppe.}
  \begin{tabular}{l l l l l l l l l l l}
    \hline
    \hline
    Transition & Freq. & $E_\mathrm{up}$ & \multicolumn{8}{c}{Fluxes} \\
    & (GHz) & (cm$^{-1}$) & \multicolumn{8}{c}{($\mathrm{erg} \, \mathrm{s}^{-1} \, \mathrm{cm}^{-2}$)}\\
    \cline{4-11}
    & & & IRAS4A & IRAS4B & IRAS2 & L1448MM & L1448N & L1157MM & L1527 & VLA1623\\
    \hline
    $2_{1,2}-1_{1,1}$     & 140.8 & 15  & 2.6E-15 & 1.6E-15 & 4.6E-16 & 1.1E-15 & 2.1E-15 & 3.7E-16 & 7.6E-16 & 3.8E-16 \\
    $2_{1,1}-1_{1,0}$     & 150.4 & 16  & 2.5E-15 & 1.4E-15 & 3.9E-16 & 9.2E-16 & 2.1E-15 & 3.5E-16 & 6.4E-16 & 3.6E-16 \\
    $3_{1,3}-2_{1,2}$     & 211.2 & 22  & 3.4E-15 & 1.6E-15 & 6.3E-16 & 1.1E-15 & 3.3E-15 & 5.6E-16 & 5.7E-16 & 4.3E-16 \\
    $3_{1,2}-2_{1,1}$     & 225.6 & 23  & 2.8E-15 & 1.1E-15 & 4.6E-16 & 6.5E-16 & 2.4E-15 & 4.4E-16 & 2.6E-16 & 3.2E-16 \\
    $4_{1,4}-3_{1,3}$     & 281.5 & 32  & 3.3E-15 & 1.1E-15 & 6.6E-16 & 7.5E-16 & 2.1E-15 & 5.3E-16 & 2.0E-16 & 3.2E-16 \\
    $4_{3,2}-3_{3,1}$     & 291.3 & 98  & 3.0E-17 & 3.3E-18 & 9.7E-18 & 4.6E-18 & 2.8E-18 & 3.7E-18 & 3.1E-19 & 1.5E-18 \\
    $4_{3,1}-3_{3,0}$     & 291.3 & 98  & 3.0E-17 & 3.3E-18 & 9.7E-18 & 4.6E-18 & 2.8E-18 & 3.7E-18 & 2.9E-19 & 1.5E-18 \\
    $4_{1,3}-3_{1,2}$     & 300.8 & 33  & 2.6E-15 & 5.8E-16 & 4.8E-16 & 4.1E-16 & 1.1E-15 & 3.9E-16 & 5.6E-17 & 2.1E-16 \\
    $5_{1,5}-4_{1,4}$     & 351.7 & 43  & 2.7E-15 & 5.3E-16 & 6.0E-16 & 4.4E-16 & 8.1E-16 & 4.0E-16 & 4.9E-17 & 2.0E-16 \\
    $5_{3,3}-4_{3,2}$     & 364.2 & 110 & 8.6E-17 & 7.1E-18 & 2.6E-17 & 9.5E-18 & 5.1E-18 & 1.0E-17 & 2.9E-19 & 3.7E-18 \\
    $5_{3,2}-4_{3,2}$     & 364.2 & 110 & 8.6E-17 & 7.1E-18 & 2.6E-17 & 9.5E-18 & 5.1E-18 & 1.0E-17 & 2.7E-19 & 3.7E-18 \\
    $5_{1,4}-4_{1,3}$     & 375.8 & 46  & 2.5E-15 & 3.8E-16 & 5.2E-16 & 3.0E-16 & 5.0E-16 & 3.6E-16 & 1.6E-17 & 1.6E-16 \\
    $6_{1,6}-5_{1,5}$     & 421.9 & 57  & 2.3E-15 & 3.1E-16 & 5.6E-16 & 2.9E-16 & 3.4E-16 & 3.2E-16 & 1.4E-17 & 1.4E-16 \\
    $6_{3,4}-5_{3,3}$     & 437.1 & 125 & 1.6E-16 & 1.0E-17 & 4.7E-17 & 1.4E-17 & 6.4E-18 & 1.8E-17 & 2.2E-19 & 5.9E-18 \\
    $6_{3,3}-5_{3,2}$     & 437.2 & 125 & 1.6E-16 & 1.0E-17 & 4.7E-17 & 1.4E-17 & 6.4E-18 & 1.8E-17 & 2.0E-19 & 6.0E-18 \\
    $6_{1,5}-5_{1,4}$     & 450.8 & 61  & 2.3E-15 & 2.3E-16 & 5.1E-16 & 2.1E-16 & 2.2E-16 & 3.0E-16 & 3.9E-18 & 1.2E-16 \\
    $7_{1,7}-6_{1,6}$     & 491.9 & 74  & 2.0E-15 & 1.9E-16 & 5.1E-16 & 1.9E-16 & 1.6E-16 & 2.7E-16 & 3.9E-18 & 1.0E-16 \\
    $7_{3,5}-6_{3,4}$     & 510.1 & 142 & 2.5E-16 & 1.2E-17 & 6.8E-17 & 1.7E-17 & 6.6E-18 & 2.7E-17 & 1.4E-19 & 7.7E-18 \\
    $7_{3,4}-6_{3,3}$     & 510.2 & 142 & 2.5E-16 & 1.2E-17 & 6.8E-17 & 1.7E-17 & 6.6E-18 & 2.7E-17 & 1.3E-19 & 7.7E-18 \\
    $7_{1,6}-6_{1,5}$     & 525.6 & 78  & 2.0E-15 & 1.5E-16 & 4.9E-16 & 1.6E-16 & 1.1E-16 & 2.6E-16 & 1.2E-18 & 9.0E-17 \\
    $8_{1,8}-7_{1,7}$     & 561.8 & 93  & 1.8E-15 & 1.2E-16 & 4.7E-16 & 1.4E-16 & 8.0E-17 & 2.2E-16 & 1.2E-18 & 7.4E-17 \\
    $8_{3,6}-7_{3,5}$     & 583.1 & 161 & 3.4E-16 & 1.3E-17 & 9.1E-17 & 1.9E-17 & 6.2E-18 & 3.6E-17 & 8.2E-20 & 8.9E-18 \\
    $8_{3,5}-7_{3,4}$     & 583.2 & 161 & 3.4E-16 & 1.3E-17 & 9.0E-17 & 1.8E-17 & 6.2E-18 & 3.6E-17 & 7.6E-20 & 8.8E-18 \\
    $8_{1,7}-7_{1,6}$     & 600.3 & 68  & 1.9E-15 & 1.1E-16 & 4.7E-16 & 1.2E-16 & 6.1E-17 & 2.3E-16 & 4.3E-19 & 6.9E-17 \\
    $9_{1,9}-8_{1,8}$     & 631.6 & 114 & 1.7E-15 & 8.7E-17 & 4.4E-16 & 1.1E-16 & 4.7E-17 & 2.0E-16 & 4.4E-19 & 5.8E-17 \\
    $9_{3,7}-8_{3,6}$     & 656.1 & 183 & 4.4E-16 & 1.4E-17 & 1.1E-16 & 1.9E-17 & 5.5E-18 & 4.5E-17 & 4.4E-20 & 9.5E-18 \\
    $9_{3,6}-8_{3,5}$     & 656.4 & 183 & 4.4E-16 & 1.4E-17 & 1.1E-16 & 1.9E-17 & 5.5E-18 & 4.4E-17 & 4.1E-20 & 9.5E-18 \\
    $9_{1,8}-8_{1,7}$     & 674.7 & 121 & 8.0E-17 & 1.8E-15 & 4.6E-16 & 9.8E-17 & 3.9E-17 & 2.1E-16 & 1.8E-19 & 5.6E-17 \\
    $10_{1,10}-9_{1,9}$   & 701.3 & 137 & 6.5E-17 & 1.7E-15 & 4.2E-16 & 8.3E-17 & 2.9E-17 & 1.9E-16 & 1.7E-19 & 4.6E-17 \\
    $10_{3,8}-9_{3,7}$    & 729.1 & 207 & 1.3E-17 & 5.3E-16 & 1.3E-16 & 1.8E-17 & 4.3E-18 & 5.1E-17 & 1.9E-20 & 9.2E-18 \\
    $10_{1,9}-9_{1,8}$    & 729.6 & 207 & 1.3E-17 & 5.2E-16 & 1.3E-16 & 1.8E-17 & 4.3E-18 & 5.1E-17 & 1.8E-20 & 9.2E-18 \\
    $11_{1,11}-10_{1,10}$ & 749.0 & 146 & 6.2E-17 & 1.8E-15 & 4.4E-16 & 7.9E-17 & 2.5E-17 & 1.9E-16 & 7.7E-20 & 4.5E-17 \\
    $11_{1,10}-10_{1,10}$ & 770.8 & 163 & 5.0E-17 & 1.7E-15 & 4.0E-16 & 6.5E-17 & 1.9E-17 & 1.7E-16 & 6.7E-20 & 3.7E-17 \\
    $11_{1,10}-10_{1,9}$  & 823.0 & 173 & 4.9E-17 & 1.8E-15 & 4.3E-16 & 6.5E-17 & 1.7E-17 & 1.9E-16 & 3.4E-20 & 3.7E-17 \\ 
    $12_{1,12}-11_{1,11}$ & 840.2 & 191 & 3.6E-17 & 1.6E-15 & 3.7E-16 & 4.8E-17 & 1.1E-17 & 1.6E-16 & 2.2E-20 & 2.8E-17 \\ 
    $12_{1,11}-11_{1,10}$ & 896.7 & 203 & 3.6E-17 & 1.8E-15 & 3.9E-16 & 4.8E-17 & 1.0E-17 & 1.7E-16 & 1.3E-20 & 2.8E-17 \\ 
    \hline    
  \end{tabular}
\label{tab:flux_const}
\end{table*}


%% file: table_flux_jump.tex
\begin{table*}
  \caption{Predicted line fluxes of selected o-H$_2$CO transitions for
    an abundance \Xout in the outer enveloppe, increasing to \Xin at
    the radius where $T_\mathrm{dust} \ge 100$ K.}
  \begin{tabular}{l l l l l l l l l l}
    \hline
    \hline
    Transition & Freq. & $E_\mathrm{up}$ & \multicolumn{7}{c}{Fluxes} \\
    & (GHz) & (cm$^{-1}$) & \multicolumn{7}{c}{($\mathrm{erg} \, \mathrm{s}^{-1} \, \mathrm{cm}^{-2}$)}\\
    \cline{4-10}
    & & & IRAS4A & IRAS4B & IRAS2 & L1448MM & L1448N & L1157MM & L1527 \\
    \hline
    $2_{1,2}-1_{1,1}$     & 140.8 & 15  & 2.7E-15 & 1.6E-15 & 4.9E-16 & 1.1E-15 & 2.1E-15 & 3.7E-16 & 7.6E-16 \\
    $2_{1,1}-1_{1,0}$     & 150.4 & 16  & 2.5E-15 & 1.4E-15 & 4.2E-16 & 9.2E-16 & 2.1E-15 & 3.6E-16 & 6.4E-16 \\
    $3_{1,3}-2_{1,2}$     & 211.2 & 22  & 3.5E-15 & 1.7E-15 & 7.3E-16 & 1.1E-15 & 3.3E-15 & 5.9E-16 & 5.7E-16 \\
    $3_{1,2}-2_{1,1}$     & 225.6 & 23  & 2.9E-15 & 1.1E-15 & 5.8E-16 & 6.9E-16 & 2.4E-15 & 4.8E-16 & 2.7E-16 \\
    $4_{1,4}-3_{1,3}$     & 281.5 & 32  & 3.5E-15 & 1.1E-15 & 8.9E-16 & 8.1E-16 & 2.2E-15 & 6.2E-16 & 2.1E-16 \\
    $4_{3,2}-3_{3,1}$     & 291.3 & 98  & 1.6E-16 & 9.7E-17 & 2.8E-16 & 6.4E-17 & 4.6E-17 & 4.5E-17 & 9.2E-19 \\
    $4_{3,1}-3_{3,0}$     & 291.3 & 98  & 1.6E-16 & 9.6E-17 & 2.8E-16 & 6.4E-17 & 4.6E-17 & 4.5E-17 & 9.0E-19 \\
    $4_{1,3}-3_{1,2}$     & 300.8 & 33  & 2.9E-15 & 6.8E-16 & 7.7E-16 & 4.9E-16 & 1.1E-15 & 5.1E-16 & 6.0E-17 \\
    $5_{1,5}-4_{1,4}$     & 351.7 & 43  & 3.1E-15 & 6.9E-16 & 1.0E-15 & 5.6E-16 & 9.2E-16 & 6.2E-16 & 5.7E-17 \\
    $5_{3,3}-4_{3,2}$     & 364.2 & 110 & 4.6E-16 & 1.9E-16 & 5.6E-16 & 1.5E-16 & 1.1E-16 & 1.5E-16 & 2.3E-18 \\
    $5_{3,2}-4_{3,2}$     & 364.2 & 110 & 4.6E-16 & 1.9E-16 & 5.6E-16 & 1.5E-16 & 1.2E-16 & 1.5E-16 & 2.3E-18 \\
    $5_{1,4}-4_{1,3}$     & 375.8 & 46  & 3.0E-15 & 5.7E-16 & 1.1E-15 & 4.5E-16 & 6.4E-16 & 6.2E-16 & 2.5E-17 \\
    $6_{1,6}-5_{1,5}$     & 421.9 & 57  & 3.0E-15 & 5.7E-16 & 1.3E-15 & 4.9E-16 & 5.4E-16 & 7.2E-16 & 2.7E-17 \\
    $6_{3,4}-5_{3,3}$     & 437.1 & 125 & 9.2E-16 & 3.2E-16 & 9.6E-16 & 2.6E-16 & 2.1E-16 & 3.2E-16 & 3.9E-18 \\
    $6_{3,3}-5_{3,2}$     & 437.2 & 125 & 9.2E-16 & 3.2E-16 & 9.6E-16 & 2.6E-16 & 2.2E-16 & 3.2E-16 & 4.0E-18 \\
    $6_{1,5}-5_{1,4}$     & 450.8 & 61  & 3.1E-15 & 5.5E-16 & 1.4E-15 & 4.7E-16 & 4.6E-16 & 7.9E-16 & 1.7E-17 \\
    $7_{1,7}-6_{1,6}$     & 491.9 & 74  & 3.1E-15 & 6.0E-16 & 1.7E-15 & 5.2E-16 & 4.6E-16 & 9.2E-16 & 2.0E-17 \\
    $7_{3,5}-6_{3,4}$     & 510.1 & 142 & 1.5E-15 & 4.9E-16 & 1.5E-15 & 3.9E-16 & 3.4E-16 & 5.7E-16 & 4.8E-18 \\
    $7_{3,4}-6_{3,3}$     & 510.2 & 142 & 1.5E-15 & 4.9E-16 & 1.4E-15 & 3.9E-16 & 3.4E-16 & 5.6E-16 & 4.9E-18 \\
    $7_{1,6}-6_{1,5}$     & 525.6 & 78  & 3.4E-15 & 6.6E-16 & 1.9E-15 & 5.6E-16 & 4.8E-16 & 1.1E-15 & 1.7E-17 \\
    $8_{1,8}-7_{1,7}$     & 561.8 & 93  & 3.4E-15 & 7.5E-16 & 2.2E-15 & 6.3E-16 & 5.4E-16 & 1.2E-15 & 1.8E-17 \\
    $8_{3,6}-7_{3,5}$     & 583.1 & 161 & 2.3E-15 & 7.4E-16 & 2.2E-15 & 5.9E-16 & 5.2E-16 & 9.1E-16 & 4.9E-18 \\
    $8_{3,5}-7_{3,4}$     & 583.2 & 161 & 2.3E-15 & 7.4E-16 & 2.2E-15 & 5.9E-16 & 5.2E-16 & 9.1E-16 & 5.0E-18 \\
    $8_{1,7}-7_{1,6}$     & 600.3 & 68  & 3.9E-15 & 8.6E-16 & 2.6E-15 & 7.2E-16 & 6.2E-16 & 1.5E-15 & 1.5E-17 \\
    $9_{1,9}-8_{1,8}$     & 631.6 & 114 & 4.1E-15 & 1.0E-15 & 3.0E-15 & 8.3E-16 & 7.2E-16 & 1.7E-15 & 1.5E-17 \\
    $9_{3,7}-8_{3,6}$     & 656.1 & 183 & 3.2E-15 & 1.0E-15 & 3.1E-15 & 8.3E-16 & 7.3E-16 & 1.3E-15 & 3.9E-18 \\
    $9_{3,6}-8_{3,5}$     & 656.4 & 183 & 3.2E-15 & 1.0E-15 & 3.1E-15 & 8.3E-16 & 7.3E-16 & 1.3E-15 & 4.0E-18 \\
    $9_{1,8}-8_{1,7}$     & 674.7 & 121 & 4.7E-15 & 1.2E-15 & 3.6E-15 & 9.6E-16 & 8.5E-16 & 2.0E-15 & 1.3E-17 \\
    $10_{1,10}-9_{1,9}$   & 701.3 & 137 & 5.1E-15 & 1.4E-15 & 4.1E-15 & 1.1E-15 & 9.8E-16 & 2.3E-15 & 1.0E-17 \\
    $10_{3,8}-9_{3,7}$    & 729.1 & 207 & 4.2E-15 & 1.4E-15 & 4.1E-15 & 1.1E-15 & 9.2E-16 & 1.8E-15 & 2.2E-18 \\
    $10_{1,9}-9_{1,8}$    & 729.6 & 207 & 4.1E-15 & 1.4E-15 & 4.1E-15 & 1.1E-15 & 9.1E-16 & 1.8E-15 & 2.2E-18 \\
    $11_{1,11}-10_{1,10}$ & 749.0 & 146 & 5.8E-15 & 1.6E-15 & 4.8E-15 & 1.3E-15 & 1.1E-15 & 2.7E-15 & 8.5E-18 \\
    $11_{1,10}-10_{1,10}$ & 770.8 & 163 & 6.3E-15 & 1.8E-15 & 5.4E-15 & 1.4E-15 & 1.3E-15 & 3.0E-15 & 6.4E-18 \\
    $11_{1,10}-10_{1,9}$  & 823.0 & 173 & 7.3E-15 & 2.1E-15 & 6.3E-15 & 1.7E-15 & 1.5E-15 & 3.5E-15 & 5.1E-18 \\ 
    $12_{1,12}-11_{1,11}$ & 840.2 & 191 & 7.9E-15 & 2.4E-15 & 7.0E-15 & 1.8E-15 & 1.6E-15 & 3.7E-15 & 2.9E-18 \\ 
    $12_{1,11}-11_{1,10}$ & 896.7 & 203 & 9.1E-15 & 2.8E-15 & 8.2E-15 & 2.1E-15 & 1.8E-15 & 4.3E-15 & 2.4E-18 \\ 
    \hline    
  \end{tabular}
  \label{tab:flux_jump}
\end{table*}
